\documentclass[11pt]{article}

\usepackage[letterpaper,margin=1in]{geometry}
\usepackage[T1]{fontenc}
\usepackage[utf8]{inputenc}
\usepackage{lmodern}
\usepackage{amsmath,amssymb,amsthm}
\usepackage{enumitem}
\usepackage{hyperref}
\usepackage{titlesec}
\usepackage{graphicx}
\graphicspath{{./}}

\titleformat{\section}{\Large\bfseries}{\thesection}{0.7em}{}
\titleformat{\subsection}{\large\bfseries}{\thesubsection}{0.7em}{}
\titleformat{\subsubsection}{\normalsize\bfseries}{\thesubsubsection}{0.7em}{}
\titlespacing*{\section}{0pt}{2.2ex plus 0.6ex minus 0.2ex}{1.0ex}
\titlespacing*{\subsection}{0pt}{1.8ex plus 0.5ex minus 0.2ex}{0.8ex}

\setlist[itemize]{topsep=0.4em,itemsep=0.25em,leftmargin=1.5em}
\setlist[enumerate]{topsep=0.4em,itemsep=0.25em,leftmargin=1.7em}

\newtheorem{theorem}{Theorem}[section]
\newtheorem{proposition}[theorem]{Proposition}
\newtheorem{lemma}[theorem]{Lemma}
\newtheorem{corollary}[theorem]{Corollary}
\newtheorem{remark}[theorem]{Remark}
\theoremstyle{definition}
\newtheorem{definition}[theorem]{Definition}

\newcommand{\E}{\mathbb{E}}
\newcommand{\1}{\mathbf{1}}
\newcommand{\veeop}{\mathbin{\vee}}
\newcommand{\cx}{\preceq_{\mathrm{cx}}}
\newcommand{\Sh}{\mathsf{S}}
\newcommand{\bolda}{\boldsymbol{a}}
\newcommand{\vqr}{\tilde{V}}
\newcommand{\Vcvx}{V^{\mathrm{cvx}}}
\newcommand{\Vaff}{V^{\mathrm{aff}}}
\newcommand{\Vrlx}{V^{\mathrm{rlx}}}
\newcommand{\Vsh}{V^{\mathrm{sh}}}

\begin{document}
\thispagestyle{empty}

\begin{center}
	{\LARGE Constructive Model-Free Lower Bound for Bermudan Options}\par\vspace{2.0em}
	{\large Timur Misirpashayev\footnote{E-mail: tmisirpasha3@bloomberg.net}}\par\vspace{0.8em}
	{\large Bloomberg}\par\vspace{0.8em}
	{\small November 23, 2026}
\end{center}

\vspace{0.6em}
\hrule
\vspace{1.0em}

\begin{abstract}
	Piterbarg \cite{Piterbarg2026Canary} established the sharp
	model-free lower bound for the Canary call, an option that can be
	exercised on either of two dates into standard European call payoffs.
	This work considers the more general lower-bound problem for a
	Bermudan option exercisable on any of $N$ dates into payoffs of no
	more than linear growth, convex from the second exercise date onward.
	The one-dimensional marginal
	distributions of the underlying are fixed at every exercise date,
	while the joint law is allowed to range over all martingale couplings.
	This gives rise to a minimax problem with
	an outer infimum over martingale couplings $Q$ and an inner supremum
	over stopping times $\tau$. We prove that this problem has no minimax
	gap and that there exist an optimal model-free stopping rule and a
	matching optimal semi-static subhedge. The sharp value of the lower
	bound is characterized as a supremum over
	convex-majorant subhedges generalizing affine straddle--butterfly
	subhedges introduced by Piterbarg for the Canary case.
	We recover sufficiency of the
	affine form for piecewise-linear convex payoff profiles
	with a single kink, such as calls or puts, at the second exercise
	date. With more than two exercise dates, the affine form still
	provides a lower bound but is not expected to be sharp in general
	because of additional kinks in the exercise profiles that develop at
	earlier exercise dates.
	Our proof involves continuous and quantized relaxations of stopping
	rules; we show that neither kind of relaxation changes the model-free
	lower value.
	Along the way, we derive general convex-optimization identities for
	expectations of convex functions over shadow measures and for
	optimization over submeasure splits.
	We also provide
	an alternative constructive expression for the lower bound in terms
	of a backward shadow recursion on submeasures.
\end{abstract}

\vspace{0.3em}
\vfill\eject

\section{Introduction}
\label{sec:introduction}
  
Model-free or robust bounds for the price of a financial derivative dependent on the values of an underlying process $X$ at multiple dates $T_1<\cdots<T_N$ are bounds that cannot be crossed under any admissible model of joint dependence $Q$ with fixed and known marginal probability measures $\mu_1,\ldots,\mu_N$ at each date.
We assume that the option pricing problem is already expressed in the form where $X$ is a martingale\footnote{We do not assume that the process $X$ is Markovian.} that we sample at discrete times, $X_i=X(T_i)$, from a joint probability law $Q$ such that $\E^Q[X_2|X_1]=X_1$, $\E^Q[X_3|X_2,X_1]= X_2$, $\ldots$, $\E^Q[X_N|X_{N-1},\ldots,X_1]=X_{N-1}$. The laws of the marginals $X_i \sim \mu_i$ are considered model-free inputs, which must be in the convex order, $\mu_1\cx\cdots\cx\mu_N$, for any martingale coupling to exist.\footnote{
	Strassen's theorem \cite{Strassen1965}.
We only consider finite measures with finite first moment; $\alpha\cx\beta$ means
$\E^{\alpha}[\varphi]\le \E^{\beta}[\varphi]$
for every convex $\varphi$ for which the expectations are defined.} The payoff at each exercise date $T_i$ is $e_i(X_i)$, assumed to be of no more than linear growth and convex\footnote{For the first payoff, convexity is not needed, and the
linear-growth assumption may be replaced by
$e_1\in L^1(\mu_1)$.} for $i\geq2$. The stopping rule is a decision to exercise the option at $T_i$ based on the observed values $X_j$ with $j\leq i$, expressed as a random variable $\tau$ taking values in the set $\{1,\ldots,N\}$ (a stopping time of the process $X_i$). The payoff of the Bermudan option under the martingale coupling $Q$ is given by 
\begin{equation*}
	V(Q)
	=
	\sup_{\tau}
	V(Q,\tau),
\end{equation*}
where
\begin{equation}
V(Q,\tau) = \E^Q[e_\tau(X_\tau)]=\E^Q\left[\sum_{i=1}^N e_i(X_i)\1_{\{\tau=i\}}\right].
\label{eq:VQtau}
\end{equation}
The object of this study is the lower bound over all admissible measures $Q$ (which are martingale couplings with given marginals),
\begin{equation}
	V^*
	=
	\inf_{Q} V(Q) 
	=
	\inf_{Q}
	\sup_{\tau}
	\E^Q[e_\tau(X_\tau)].
	\label{eq:V}
\end{equation}
This can be regarded as an adversarial game with one player trying to maximize $V(Q,\tau)$ by choosing a stopping rule $\tau$ and the other trying to minimize $V(Q,\tau)$ by adjusting the martingale coupling $Q$. 

The main result of this work is an explicit expression for $V^*$ in terms of an optimal semi-static subhedge over $N-1$ convex majorants.\footnote{
	We also provide an alternative constructive representation for $V^*$
	in terms of a backward recursion, introduced after
	Definition~\ref{def:vshadow}.
} The space of admissible majorants $R_i$ consists of convex functions of no more than linear growth
satisfying a sequence of dominating constraints.
 In the case of the Canary option, there is a single majorant $R$ and the expression for the lower bound is
\begin{equation}
	V^* = \sup_{R\geq e_2} \left\{\E^{\mu_1}[e_1 \veeop R] + \E^{\mu_2}[e_2 - R]\right\}.
	\label{eq:V2}
\end{equation}
In the general case,
\begin{equation}
	V^* = \sup_{\substack{R_1 \geq e_2 \veeop R_2\\ \vdots \\ R_{N-2} \geq e_{N-1}\veeop R_{N-1}\\ R_{N-1}\geq e_N}} \left\{\E^{\mu_1}[e_1\veeop R_1] + \sum_{i=2}^{N-1} \E^{\mu_{i}}[e_{i}\veeop R_i - R_{i-1}] + \E^{\mu_N}[e_N-R_{N-1}]\right\}.
	\label{eq:VN}
\end{equation}

In the case of the Canary option with a call payoff at the second date, $e_2(X)=s_2(X-k_2)^+$, considered by Piterbarg \cite{Piterbarg2026Canary}, the optimization in Eq.~(\ref{eq:V2}) can be reduced from all convex majorants to those of a special form $R = e_2 \veeop \ell$ where $\ell(x) = a+bx$ is an affine function. We show that this reduction goes through for this and other terminal payoffs with a single kink. 

As a corollary of representation~(\ref{eq:VN}), we will obtain the $Q$-$\tau$ duality,
\begin{equation}
	\inf_{Q}
	\sup_{\tau}
	\E^Q[e_\tau(X_\tau)]
	=
	\sup_{\tau}
	\inf_{Q}
	\E^Q[e_\tau(X_\tau)],
	\label{eq:NoMinMaxGap}
\end{equation}
and the existence of a saddle pair $\tau^*$, $Q^*$ such that
\begin{equation}
V(Q^*,\tau)	\leq V(Q^*,\tau^*) = V^* \leq V(Q,\tau^*),
\end{equation}
for any admissible $Q$ and $\tau$.  
The stopping rule $\tau^*$ leads to an expected payoff of at least
$V^*$ for every admissible martingale coupling $Q$. Conversely,
under the minimizing coupling $Q^*$, the expected payoff does not
exceed $V^*$ for any stopping rule $\tau$.
The accompanying semi-static subhedge provides a pathwise
guarantee: its payoff is dominated by the option payoff under
$\tau^*$ on every path, and the two payoffs agree
$Q^*$-almost surely.

The attainment of the infimum over measures in the left-hand side of Eq.~(\ref{eq:NoMinMaxGap}) is a technical consequence of compactness and semicontinuity in 1-Wasserstein metric.\footnote{See Appendix~\ref{sec:terminology} for the definition.} The attainment of the supremum over stopping rules in the right-hand side cannot be obtained from topological considerations due to absence of a suitable notion of compactness. We will obtain it as a corollary of the representation~(\ref{eq:VN}).
Indeed, we will show that if the supremum is attained with majorants $R_1^*,\ldots,R_{N-1}^*$, an optimal stopping rule is ``exercise on $T_i$ where $i$ is the smallest index such that $e_i(X_i)\geq R_i^*(X_i)$, hold to $T_N$ if there is no such $i$''. The proof is then completed by showing that an optimal exercise rule also exists if the supremum is not attained, which may happen if some of the majorants escape to infinity.

In contrast with many of the works on model-free hedging,  we are able to bypass explicit use of the duality of measures and hedges with the attendant issues of duality gap and extensive algebraic manipulations. This is because we have at our disposal the shadow measure, a powerful optimization tool originated by  Beiglb\"ock and Juillet \cite{BeiglbockJuillet2016,BeiglbockJuillet2021} with an explicit potential formula due to Beiglb\"ock et al. \cite{BeiglbockHobsonNorgilas2022}, which reduces the one-step minimization over measures to calculations of put options prices and lower convex envelopes.\footnote{We became aware after this work had been completed that the
substantially revised version of \cite{Piterbarg2026Canary}
independently uses shadow measures and stopping rule relaxation to prove the sharpness of the affine bound for the two-date Canary-call
problem.}
We use in an essential manner the discrete version of shadow measure associativity. A more involved continuous version, known as the lifted shadow coupling theorem of \cite{BeiglbockJuillet2021}, is avoided although it could have been used to provide a more straightforward alternative in one of the steps. Another prominent technique that we employ is the relaxation of stopping rules in the form of liquidation strategies by Bayraktar and Zhou \cite{BayraktarZhou2016}. A more detailed discussion of connections with the prior work is deferred until Sect.~\ref{sec:discussion_directions}.

The paper is organized as follows. Sect.~\ref{sect:one-step} revisits the Canary option.  We define the convex-majorant bound in Sect.~\ref{sect:canary_bounds} and show that it is a lower bound from a semi-static hedging argument. The rest of Sect.~\ref{sect:one-step} consists of several steps proving  the convex-majorant bound is sharp. We review the shadow measure construction and show that it provides a minimizer over measures in Sect.~\ref{sect:shadow}. Sect.~\ref{sect:randomized_canary} introduces relaxed stopping rules as a tool to satisfy the conditions of a standard minimax theorem. Sect.~\ref{sect:purify} justifies the return to ordinary stopping rules by an argument involving shadow measure associativity. After that, we show with the aid of the shadow potential formula how the evaluation of expectations over shadow measures leads to the convex-majorant bound reappearing as an upper bound, first for European option payoffs in Sect.~\ref{sec:two-date-affine} and then for arbitrary convex payoffs in Sect.~\ref{sect:gctp}. This completes the proof of Eq.~(\ref{eq:V2}) with Piterbarg's affine bound as a particular case. 

We proceed to the general case of more than two exercise dates in Sect.~\ref{sec:N-majorant}, adapting the steps from Sect.~\ref{sect:one-step} and overcoming additional technical complications. We begin 
with the definition of the convex-majorant bound and a semi-static hedging argument showing that  it is a lower bound in Sect.~\ref{sec:subhedgeN}. Two kinds of relaxed stopping rules, continuous and quantized, are defined in Sect.~\ref{sec:relaxed_rulesN}. Inf-sup exchange is only enabled under continuous relaxed rules, so we will need additional reasoning to justify eventual return to ordinary stopping rules via relaxed quantized rules. 
This requires an auxiliary upper bound constructed in Sect.~\ref{sec:recursion-first} with the aid of a backward induction on submeasures. We are finally able to identify this upper bound with the convex-majorant bound and summarize the results as a main theorem in Sect.~\ref{sec:W-to-majorants}. 

With the results of Sect.~\ref{sec:N-majorant}, we established that the infimum over measures and the supremum over stopping rules in the problem of model-free Bermudan pricing can be interchanged and that the model-free lower bound can be found as a supremum over convex majorants. We could not yet claim the existence of a saddle point in this minimax problem. This question of attainability is settled positively in Sect.~\ref{sec:majorant-attainment}. In Sect.~\ref{sec:litrev}, we summarize the position of our work and technical tools that we used in the context of martingale optimal transport literature. In Sect.~\ref{sec:directions}, we consider directions for future development, including extensions to the case of continuous exercise (American options), more general payoffs, and numerical implementation for practical use. We conclude in Sect.~\ref{sec:conclusions}. Appendix~\ref{sec:terminology} provides a summary of terminology, notations, and assumptions. Appendix~\ref{sect:examples_shadow} shows two simple examples of shadow measure construction.

\section{Two exercise dates: Canary option}
\label{sect:one-step}

We begin with the case of two exercise dates, which serves as a prototype for
the subsequent multiperiod construction. 
We will derive Eq.~(\ref{eq:V2}) and recover Piterbarg's expression for the lower bound as an optimizer over straddle--butterfly hedges for the Canary call payoff.
Our formulation permits an
arbitrary integrable first-date payoff and an arbitrary convex
terminal payoff of no more than linear growth and does not assume at any point that the marginals have finite support.

The exercise rule can be conveniently expressed as a continuation region $H$, a Borel set, such that the option is exercised on $T_2$ if $X_1\in H$. The option is exercised on $T_1$ if $X_1 \in H^c$, so that the payoff under a particular martingale coupling $Q$ is given by
\begin{equation}
	V(Q,H)
	=
	\E^Q\!\left[\1_{H^c}(X_1)e_1(X_1)+\1_H(X_1)e_2(X_2)\right].
	\label{eq:VQH-intro}
\end{equation}
The first term in the right-hand side corresponds to an early exercise; the second is the continuation value. We assume that the payoff $e_2$ is convex, while no convexity assumption on $e_1$ is needed.

\subsection{Convex and affine bounds for the Canary option from semi-static subhedge}
\label{sect:canary_bounds}
The term corresponding to the early exercise depends only on $\mu_1$ but not on the coupling,
 \begin{equation}
 	V(Q,H)
 	=
 	\E^{\mu_1}\!\left[\1_{H^c}(X_1)e_1(X_1)\right]
 	+
 	\E^Q\!\left[\1_H(X_1)e_2(X_2)\right].
 	\label{eq:VQH}
 \end{equation}
We aim to construct a subhedge depending only on $\mu_1$ and $\mu_2$ for the term corresponding to the continuation value. 
Let $R\geq e_2$ be a convex function of no more than linear growth. For every $x$, we can take 
a measurable subgradient $\Delta(x)$, so that for any $x_1,x_2$ 
\begin{equation}
R(x_2)-R(x_1)\geq \Delta(x_1)(x_2-x_1).
\label{eq:ConvR}
\end{equation}
We set up a delta subhedge,
\begin{equation}
\1_H(x_1)R(x_1)+e_2(x_2)-R(x_2) +\1_H(x_1)\Delta(x_1)(x_2-x_1)
\le
\1_H(x_1)e_2(x_2).
\label{eq:one-step-pathwise}
\end{equation}
Indeed, if $x_1\in H$, this reduces to inequality~(\ref{eq:ConvR}), while for $x_1\in H^c$ we get $e_2(x_2)-R(x_2)\leq 0$.
Taking the expectation over any martingale coupling $Q$ eliminates the term with $x_2-x_1$, leaving us with a sum of payoffs dependent only on the marginals $\mu_1$, $\mu_2$ in the left-hand side. Adding back the term corresponding to the early exercise we get 
\begin{equation*}
	\E^{\mu_1}[\1_{H^c}e_1 + \1_H R]+\E^{\mu_2}[e_2-R]
	\leq
	V(Q,H).
\end{equation*}
Taking the supremum over $H$ simplifies the first term in the left-hand side,
\begin{equation*}
	\E^{\mu_1}[e_1\veeop R]+\E^{\mu_2}[e_2-R] \leq \sup_H V(Q,H),
\end{equation*}
and leads to the following convex-majorant bound on the model-free value of the Canary option
\begin{equation}
\Vcvx = \sup_{R\geq e_2} \E^{\mu_1}[e_1\veeop R]+\E^{\mu_2}[e_2-R] \leq \inf_Q\sup_H V(Q,H) = V^*.
\label{eq:lcvx2}
\end{equation} 

Piterbarg's optimization over straddle--butterfly subhedges  is obtained by restricting the class of convex majorants to the form
\begin{equation*}
R=e_2\veeop\ell,
\qquad
\ell(x)=c+\theta x,
\end{equation*}
leading to an affine bound\footnote{
	Piterbarg wrote the affine bound as $\E^{\mu_1}[(e_1 \veeop e_2 -\ell)^+] + \E^{\mu_2}[e_2 \wedge \ell]$, which is the same as Eq.~(\ref{eq:Laffine}) because 
	$
	e_1\vee e_2\vee\ell
	=
	(e_1\veeop e_2-\ell)^+ +\ell$,
	$e_2-e_2\vee\ell
	=
	e_2\wedge \ell-\ell$, and $\E^{\mu_1}[\ell]=\E^{\mu_2}[\ell]$.} 
\begin{equation}
	\Vaff = \sup_{\ell}\left\{ \E^{\mu_1}[e_1\veeop e_2\veeop \ell]+\E^{\mu_2}[e_2-e_2\veeop \ell]\right\}.
	\label{eq:Laffine}
\end{equation} 
So far, we have
\begin{equation*}
	\Vaff \leq \Vcvx \leq V^*.
\end{equation*}
We will ultimately show that $\Vcvx = V^*$ holds without additional assumptions and that $\Vaff = V^*$ under a restriction on the payoff $e_2$, satisfied in particular by the European call $e_2(x) = s_2(x-k_2)^+$.

\subsection{A fixed continuation set is minimized by a shadow}
\label{sect:shadow}

Consider an auxiliary problem of minimization of the continuation value over measures $Q$ for a fixed continuation set $H$. 
We show that the minimum is attained and that its value is given by
an expectation over the shadow measure introduced by Beiglb\"ock and
Juillet \cite{BeiglbockJuillet2016}.
\begin{lemma}[Minimizing conditional convex payoff over measures with a shadow]
	\label{lemma:minQbyshadow}
	For a convex payoff $e_2$ growing no more than linearly and a Borel
	continuation set $H$, the minimum over martingale couplings $Q$ with
	marginals $\mu_1,\mu_2$ is attained and is equal to
	\begin{equation}
	\min_Q \E^Q\!\left[\1_H(X_1)e_2(X_2)\right]
	=
	\E^{\Sh^{\mu_2}(\mu_1|_H)}[e_2].
	\label{eq:ShMin}
	\end{equation}
\end{lemma}
Before proceeding to the proof, we recall the definition of the shadow measure. Let 
$\alpha$ and $\beta$ be finite measures (not necessarily normalized to 1).
Assume that there is at least one measure $\eta$ such that $\alpha \cx \eta$ (convex order) and $\eta \leq \beta$ (order on measures as set functions, $\eta(B)\leq \beta(B)$ for any Borel set $B$). The shadow measure $\Sh^\beta(\alpha)$
is the least in convex order of all measures that satisfy the two properties of $\eta$ stated above. In view of Strassen's theorem, there exists a martingale coupling between $\alpha$ and $\Sh^\beta(\alpha)$. Informally, $\Sh^\beta(\alpha)$ is the least spread out distribution that can be obtained from $\alpha$ in a martingale transport step without putting any mass outside $\beta$. 
 
Beiglb\"ock et al. \cite{BeiglbockHobsonNorgilas2022} elevated the shadow measure from an abstract construct to a remarkably practical tool with an explicit potential formula. Associated with a finite measure $\xi$ with finite first moment is a potential $P_\xi(k) = \E^\xi[(k-x)^+]$,
known in financial applications as the put option price as a function of strike. According to Theorem 4.7 in \cite{BeiglbockHobsonNorgilas2022},
\begin{equation}
	P_{\Sh^\beta(\alpha)} = P_\beta - (P_\beta - P_\alpha)^c,
	\label{eq:shadowP}
\end{equation}
where $f^c$ is the greatest convex minorant of $f$, also called its
lower convex envelope, which is the largest convex function dominated by $f$.
We refer to \cite{BeiglbockJuillet2016}, \cite{BeiglbockHobsonNorgilas2022} for the proofs and discussion. 

The specific context in which shadow measures appear in our work involves two probability measures $\mu_1 \cx \mu_2$ and submeasures of $\mu_1$, that is, measures of the form $q(x)\mu_1$ with Radon--Nikodym derivative $0\leq q(x)\leq1$. This class includes restrictions of $\mu_1$ on Borel subsets $H \subseteq \mathbb{R}$. We can take $\alpha=\mu_1|_H$, $\beta=\mu_2$ and obtain a shadow measure $\Sh^{\mu_2}(\mu_1|_H)$ which is a submeasure of $\mu_2$.
The shadow measures are associative,\footnote{
Theorem 4.8 in \cite{BeiglbockHobsonNorgilas2022}.
Note that $H_1$ and $H_2$ do not enter symmetrically in Eq.~(\ref{eq:shadow_assoc}). After taking the shadow of $\mu_1$ in $\mu_2$, we only use the remainder $\mu_2-\Sh^{\mu_2}(\mu_1|_{H_1})$ as the target for $\mu_1|_{H_2}$. The split of $\Sh^{\mu_2}(\mu_1|_{H_1\sqcup H_2})$ would be different if we took $\Sh^{\mu_2}(\mu_1|_{H_2})$ first and $\Sh^{\mu_2-\Sh^{\mu_2}(\mu_1|_{H_2})}(\mu_1|_{H_1})$ second but the total would remain the same. Note also that generally $\Sh^{\mu_2}(\mu_1|_{H_1}) + \Sh^{\mu_2}(\mu_1|_{H_2})\neq \Sh^{\mu_2}(\mu_1|_{H_1\sqcup H_2})$, so we speak of associativity rather than additivity of shadows.
} in the sense that for disjoint sets $H_1$, $H_2$
\begin{equation}
	\Sh^{\mu_2}(\mu_1|_{H_1}) + \Sh^{\mu_2-\Sh^{\mu_2}(\mu_1|_{H_1})}(\mu_1|_{H_2}) = \Sh^{\mu_2}(\mu_1|_{H_1\sqcup H_2}).
	\label{eq:shadow_assoc}
\end{equation}
In particular, 
\begin{equation}
	\Sh^{\mu_2}(\mu_1|_{H}) + \Sh^{\mu_2-\Sh^{\mu_2}(\mu_1|_{H})}(\mu_1|_{H^c}) = \mu_2.
	\label{eq:shadowHHc_assoc}
\end{equation}
Examples of shadow measures are in Appendix~\ref{sect:examples_shadow}.

\begin{proof}
We now prove Eq.~(\ref{eq:ShMin}).
Restricting the first dimension of $Q$ to $H$, the left-hand side is the expectation under the measure $\nu_{Q,H}$ defined by $\nu_{Q,H}(B) = Q(H\times B)$ for a Borel set $B$. Obviously, $\nu_{Q,H}\leq \mu_2$. We will also show that the restriction $\mu_1|_H$ and $\nu_{Q,H}$ are in convex order. Indeed, for any convex 
$\varphi$ such that the expectations exist,
\begin{align}
\E^{\nu_{Q,H}}[\varphi]
=
\E^{Q}[\mathbf{1}_H(X_1)\varphi(X_2)]
&= 
\E^{Q}[\mathbf{1}_H(X_1)\E^{Q}[\varphi(X_2)\mid X_1]] \nonumber \\
&\ge
\E^{Q}[\mathbf{1}_H(X_1)\varphi(\E^{Q}[X_2\mid X_1])]
=
\E^{\mu_1|_H}[\varphi].
\end{align}
By minimality of the shadow, $\Sh^{\mu_2}(\mu_1|_H) \cx \nu_{Q,H}$, in particular, since $e_2$ is convex,
\begin{equation*}
\E^{\Sh^{\mu_2}(\mu_1|_H)}[e_2]
\le
\E^{\nu_{Q,H}}[e_2].
\end{equation*}
To complete the proof, we need to show that the lower bound is attainable, that is, the existence of an extension of 
a martingale coupling between $\mu_1|_H$ and $\Sh^{\mu_2}(\mu_1|_H)$ to a coupling of $\mu_1$ and $\mu_2$. 
This follows from the associativity~(\ref{eq:shadowHHc_assoc}).
Strassen's theorem provides one martingale transport from $\mu_1|_H$ to $\Sh^{\mu_2}(\mu_1|_H)$ and another from $\mu_1|_{H^c}$ to the residual target $\mu_2-\Sh^{\mu_2}(\mu_1|_H)$, adding up to a full martingale coupling of $\mu_1$ and $\mu_2$.
\end{proof}

\subsection{Randomized continuation and minimax}
\label{sect:randomized_canary}
In view of the previous subsection, it would be advantageous to exchange the order of optimizations and do the minimization over measures $Q$ first. Unfortunately, we can only immediately claim an elementary minimax inequality,
\begin{equation*}
	\sup_H \inf_Q V(Q,H) \leq \inf_Q \sup_H V(Q,H).
\end{equation*}
If the inequality happened to be strict then we would not be able to obtain a sharp lower bound as $\sup_H \inf_Q V(Q,H)$. We will show that the minimax gap is in fact absent.

We would like to use a version of Sion's minimax theorem \cite{Sion1958}, which is not immediately possible because the optimization over continuation sets $H$ lacks a necessary notion of convexity. We follow the route of Bayraktar and Zhou \cite{BayraktarZhou2016}, passing from the space 
of indicators $\1_H$ to its convex hull, $\sum_i \lambda_i \1_{H_i}$ ($0<\lambda_i\leq1$, $\sum_i \lambda_i = 1$).  With that in mind, we introduce a relaxed optimization problem on the functional
\begin{equation}
	\vqr(Q,a)
	=
	\E^Q\!\left[(1-a(X_1))e_1(X_1)+a(X_1)e_2(X_2)\right],
	\label{eq:FQa}
\end{equation}
where $a(x) : \mathbb{R} \to [0,1]$ is a Borel function taking a finite number of values. Note that $\vqr(Q,\1_{H})=V(Q,H)$. 

This relaxation of the stopping rule can alternatively be viewed as 
randomization of the exercise decision  
(a pathwise value of $X_1$  does not determine whether to exercise or continue, but only the probability $a(X_1)$ of continuation) or as a liquidation strategy (a fraction of the option is exercised on $T_1$ and the rest at $T_2$). In contrast with \cite{BayraktarZhou2016}, we do not consider nonlinear functionals of the relaxed payoff, so we are free to choose either interpretation because both lead to Eq.~(\ref{eq:FQa}).
We will speak of relaxed or randomized stopping rules as synonyms, and we will sometimes refer to ordinary stopping rules as pure.   

The randomization of exercise does not change the option value under a fixed measure $Q$, 
\begin{equation}
	\sup_a \vqr(Q,a) = \sup_H V(Q,H),
	\label{eq:supAsupH}
\end{equation}
because there is always an optimal exercise region determined by a pointwise comparison of $e_1(X_1)$ and $\E^Q[e_2(X_2)|X_1]$, which dominates any relaxed exercise rule.
 
   The relaxed minimax problem $\inf_Q \sup_a \vqr(Q,a)$ satisfies the conditions of Sion's minimax theorem. Indeed, the space of admissible measures $Q$ is convex and compact in the $1$-Wasserstein metric, the set of admissible $a$ is convex, the functional $\vqr(Q,a)$ is affine in both arguments and continuous\footnote{
Continuity in $Q$ holds even when $a$ is only bounded and Borel.
Indeed, approximate $a$ in $L^1(\mu_1)$ by bounded continuous
functions $a_n$.  Uniformly over all admissible couplings $Q$,
\begin{equation*}
\left|\E^Q[(a-a_n)(X_1)e_2(X_2)]\right|
\leq
K\|a-a_n\|_{L^1(\mu_1)}
+
\int_{\{|e_2|>K\}}|e_2|\,d\mu_2.
\end{equation*}
The right-hand side tends to zero by first letting $n\to\infty$ and
then $K\to\infty$.
} in $Q$ (in the $1$-Wasserstein metric) as well as $a$ (in the supremum norm). Therefore,
\begin{equation}
\inf_Q\sup_a \vqr(Q,a) = \sup_a\inf_Q \vqr(Q,a).
	\label{eq:sion}
\end{equation}
Combining with Eq.~(\ref{eq:supAsupH}), we get
\begin{equation}
	\inf_Q\sup_H V(Q,H) = \sup_a\inf_Q \vqr(Q,a).
\label{eq:supAinfQ}
\end{equation}

\subsection{Return to ordinary stopping rules}
\label{sect:purify}

\begin{proposition}
\begin{equation}
	\sup_a\inf_Q \vqr(Q,a) = \sup_H\inf_Q V(Q,H).
	\label{eq:supHinfQ}
\end{equation}
\end{proposition}
\begin{proof}
This will be facilitated by our ability to compute the infimum over $Q$ with shadows using Lemma~\ref{lemma:minQbyshadow}.
Indeed, any relaxed rule $a$ is a finite linear combination $\sum_{i=1}^M \lambda_i \1_{H_i}$ of Borel set indicators $\1_{H_i}$ where the sets $H_i$ can be chosen disjoint. 
We can assume without loss of generality\footnote{
We can add $\1_\varnothing$ to $a$ if necessary to ensure that $\lambda_M=1$. 
} that $0<\lambda_1<\cdots<\lambda_M = 1$ and switch to the ``layer cake'' representation for the relaxed stopping rule $a$, 
\begin{equation}
	a = \sum_{i=1}^M (\lambda_i-\lambda_{i-1})\1_{U_i}, \qquad U_i = \sqcup_{j=i}^M H_j, \qquad \lambda_0=0.
\end{equation}

 The first term in Eq.~(\ref{eq:FQa}) does not depend on $Q$. The minimum of the second term can be evaluated in terms of shadow measures,  
\begin{align*}
	\min_Q \E^Q[a(X_1)e_2(X_2)] &= \sum_{i=1}^M (\lambda_i-\lambda_{i-1}) \min_Q \E^Q[\1_{U_i}(X_1)e_2(X_2)]
	\nonumber \\ &= \sum_{i=1}^M (\lambda_i-\lambda_{i-1}) \E^{\Sh^{\mu_2}(\mu_1|_{U_i})}[e_2].
\end{align*}
It follows from repeated application of associativity of the shadows~(\ref{eq:shadow_assoc}) that for a sequence of nested sets, $U_1\supseteq\cdots\supseteq U_M$ a common martingale coupling can be chosen for all pairs of $\mu_1|_{U_i}$ and $\Sh^{\mu_2}(\mu_1|_{U_i})$.
Thus the following equality holds, and all terms in the right-hand side are simultaneously minimized by a common measure $Q=Q^a$.
\begin{equation*}
	\min_Q \vqr(Q,a) =  \sum_{i=1}^M (\lambda_i-\lambda_{i-1}) \min_Q V(Q,U_i).
\end{equation*}

Now
\begin{equation*}
	\sum_{i=1}^M (\lambda_i-\lambda_{i-1}) \min_Q V(Q,U_i) \leq \max_i \min_Q V(Q,U_i),
\end{equation*}
because the weights sum up to $\lambda_M=1$.
A maximum over sets $U_i$ is not greater than the supremum over all Borel sets $H$, therefore,
\begin{equation*}
	\min_Q \vqr(Q,a) \leq \sup_H \min_Q V(Q,H).
\end{equation*}
Taking the supremum over $a$, 
\begin{equation*}
	\sup_a \min_Q \vqr(Q,a) \leq \sup_H \min_Q V(Q,H).
\end{equation*}
The opposite inequality is obvious since $\1_{H}$ can be regarded as a relaxed rule. Therefore, the equality holds, which completes the proof of Eq.~(\ref{eq:supHinfQ}).
\end{proof}
The absence of minimax gap for the original problem follows. Furthermore, the lower-bound problem takes the form,
\begin{equation}
	V^*=\sup_H \inf_Q V(Q,H) =\sup_H \left\{\E^{\mu_1|_{H^c}}[e_1] + \E^{\Sh^{\mu_2}(\mu_1|_{H})}[e_2]\right\}.
	\label{eq:supinfV}
\end{equation}

\subsection{European call at the second exercise}
\label{sec:two-date-affine}

The potential formula~(\ref{eq:shadowP}) in the form given in \cite{BeiglbockHobsonNorgilas2022} directly gives put prices under shadow measure, which is particularly convenient if the payoff $e_2$ is that of a put option. If we want to handle first the case of a call payoff, $e_2(x) = s(x-k)^+$, considered by Piterbarg, it is more convenient to switch to a representation in terms of call options,\footnote{
Eq.~(\ref{eq:shadowC}) follows from Eq.~(\ref{eq:shadowP}), call-put parity, $C^\xi(k)-P^\xi(k) = \E^\xi[x-k]$, possibility of taking any affine function $\ell$ outside of lower convex envelope, $(f+\ell)^c = f^c+\ell$, and conservation of mass and first moment for a shadow, $\E^{\Sh^\beta(\alpha)}[x-k] = \E^\alpha[x-k]$.
} $C_\xi(k) = \E^\xi[(x-k)^+]$,
\begin{equation}
		C_{\Sh^\beta(\alpha)} = C_\beta - (C_\beta - C_\alpha)^c.
	\label{eq:shadowC}
\end{equation} 
We have
\begin{equation}
	\E^{\Sh^{\mu_2}(\mu_1|_{H})}[e_2] =s\left(C_{\mu_2}(k) -(C_{\mu_2}-C_{\mu_1|_H})^c(k)\right).
	\label{eq:sh_e2}
\end{equation}
The value of the lower convex envelope $f^c$ at any point $k$ can be obtained as the infimum of values of all chords connecting the points $(l,f(l))$ and $(r,f(r))$ with $l\leq k \leq r$, $l<r$, 
\begin{equation}
f^c(k)
=
\inf_{\substack{l\le k\le r \\ l < r}}
\left\{
pf(l)+qf(r)
\right\},
\label{eq:chord-formula}
\end{equation}
\begin{equation}
p = \frac{r-k}{r-l},\qquad
q = \frac{k-l}{r-l}.
\end{equation}
Define
\begin{equation}
R_{l,r}(x) = 
s(p(x-l)^++q(x-r)^+), \qquad \ell_{l,r}(x)=sp(x-l),
\label{eq:split-hinge}
\end{equation}
then $R_{l,r} = e_2 \veeop \ell_{l,r}$ (geometrically, the line $\ell_{l,r}$ cuts over the corner of the payoff profile $e_2(x)=s(x-k)^+$, providing an affine lift on $[l,r]$).
Substituting Eq.~(\ref{eq:chord-formula}) into Eq.~(\ref{eq:sh_e2}), we get after rearrangements
\begin{equation}
	\E^{\Sh^{\mu_2}(\mu_1|_{H})}[e_2] = \sup_{\substack{l\le k\le r \\ l < r}}\left\{\E^{\mu_1|_H}[R_{l,r}] + \E^{\mu_2}[e_2 - R_{l,r}]\right\}. 
	\label{eq:e_sh_rl}
\end{equation}

We now plug Eq.~(\ref{eq:e_sh_rl}) into the full stopping problem~(\ref{eq:supinfV}), getting
\begin{equation*}
	V^* = \sup_H\inf_Q V(Q,H) = \sup_H \sup_{\substack{l\le k\le r \\ l < r}}\left\{\E^{\mu_1}[\1_{H^c}e_1 + \1_H R_{l,r}] + \E^{\mu_2}[e_2 - R_{l,r}]
	 \right\}.
\end{equation*}
Interchanging the order, the supremum over $H$ is simply $\sup_H \E^{\mu_1}[\1_{H^c}e_1 + \1_H R_{l,r}] = \E^{\mu_1}[e_1\veeop R_{l,r}]$, so
\begin{equation*}
	V^* =\sup_{\substack{l\le k\le r \\ l < r}}\left\{\E^{\mu_1}[e_1\veeop R_{l,r}] + \E^{\mu_2}[e_2 - R_{l,r}]\right\}.
\end{equation*}
Recalling that $R_{l,r}=e_2\veeop \ell_{l,r}$, we have nearly obtained Piterbarg's affine bound $\Vaff$ given by Eq.~(\ref{eq:Laffine}).
The only difference is that in 
\begin{equation*}
	V^* =\sup_{\substack{l\le k\le r \\ l < r}}\left\{\E^{\mu_1}[e_1\veeop e_2 \veeop \ell_{l,r}] + \E^{\mu_2}[e_2 - e_2\veeop \ell_{l,r}]\right\},
\end{equation*}
we do not have a supremum over all affine functions but only over a subset. Extending the optimization to all affine functions might increase the supremum, so $V^*\leq \Vaff$. However, we already proved that $\Vaff \leq V^*$, so $V^*=\Vaff$, and we recover Piterbarg's lower bound for the Canary option \cite{Piterbarg2026Canary}.

\subsection{General convex terminal payoffs}
\label{sect:gctp}
We proceed to extend the result to more general convex payoffs $e_2$. A straightforward extension is to a piecewise-linear payoff with a single kink, which is the call plus an affine function, $e_2(x)=a+bx+s(x-k)^+$. This, in particular, includes the put payoff, $s(k-x)^+ = s(k-x) + s(x-k)^+$.  Due to martingality, an affine addend $\ell_0=a+bx$ to $e_2$ can be subsumed into $e_1$,
\begin{equation}
	\E^{\mu_1}[\1_{H^c}e_1 ] + \E^Q[\1_H(X_1)(e_2(X_2)+\ell_0(X_2))] = \E^{\mu_1}[\ell_0] + \E^{\mu_1}[\1_{H^c}(e_1-\ell_0) ] + \E^Q[\1_H(X_1)e_2(X_2)], 
\end{equation}
then we can write the affine bound for the payoffs $e_1-\ell_0$, $e_2$ as a supremum over lines,  
and after rearrangements we end up with the same form~(\ref{eq:Laffine}) of the affine bound for the payoffs $e_1$, $e_2+\ell_0$.
(An even simpler way of extending the affine bound to the put terminal payoff is, of course, reflecting the process $X_i\to -X_i$, which turns puts into calls with reflected strikes.)

We should not expect the affine form to hold generally for a convex terminal payoff $e_2$ with more than one kink or continuously distributed curvature. Consider a payoff with several kinks,
\begin{equation}
e_2(x)=\sum_m s_m(x-k_m)^+,
\qquad
s_m>0.
\label{eq:PWL-C}
\end{equation}
In evaluating the expectation over the shadow measure, we will get
a sum of several terms similar to the right-hand side of Eq.~(\ref{eq:sh_e2}) but corresponding to different strikes $k_m$. In each of them, the operation of taking the lower convex envelope can be replaced by an optimizer over lines connecting the points on the graph with abscissas $l_m<r_m$ such that $l_m\leq k_m \leq r_m$. The result is 
\begin{equation}
	\E^{\Sh^{\mu_2}(\mu_1|_{H})}[e_2] = \sup_{\substack{l_m\le k_m\le r_m \\ l_m < r_m}}\left\{\E^{\mu_1|_H}[R_{\mathbf{l},\mathbf{r}}] + \E^{\mu_2}[e_2 - R_{\mathbf{l},\mathbf{r}}]\right\}, 
	\label{eq:e_sh_rl_vec}
\end{equation}
where
\begin{equation}
R_{\mathbf{l},\mathbf{r}}(x)
=
\sum_m s_m \left(p_m(x-l_m)^++q_m(x-r_m)^+\right),\qquad p_m=\frac{r_m-k_m}{r_m-l_m}, \quad q_m=\frac{k_m-l_m}{r_m-l_m}.
\label{eq:bridge-majorant}
\end{equation}
Obviously, $R_{\mathbf{l},\mathbf{r}}$ is convex and of no more than linear growth and $R_{\mathbf{l},\mathbf{r}}\geq e_2$.
The next step, where the order of maximizations over continuation sets $H$ and convex majorants $R_{\mathbf{l},\mathbf{r}}$ is interchanged and the maximization over $H$ is done, goes through unchanged, and we end up with the following expression for the lower bound
\begin{equation}	
V^* = \sup_{\substack{l_m\le k_m\le r_m \\ l_m < r_m}}\left\{\E^{\mu_1}[e_1\veeop R_{\mathbf{l},\mathbf{r}} ]+ \E^{\mu_2}[e_2 - R_{\mathbf{l},\mathbf{r}} ]\right\}.
\end{equation}
Expanding the optimization to a larger set of all convex functions $R\geq e_2$ of no more than linear growth may turn the equality into the inequality $V^* \leq \Vcvx$ with $\Vcvx$ defined by (\ref{eq:lcvx2}). As we already proved that $V^* \geq \Vcvx$, we conclude that $V^*=\Vcvx$, so the convex-majorant bound is sharp for the Canary terminal payoff of the form~(\ref{eq:PWL-C}).

Similarly to the case of a single kink, the addition of a linear term
to $e_2$ does not alter the form of the convex-majorant bound.  Furthermore, any
convex payoff $e_2$ of no more than linear growth can be approximated
from above, locally uniformly, by convex piecewise-linear functions of
the form
\begin{equation*}
	a_n+b_nx+\sum_{m=1}^{M_n}s_{n,m}(x-k_{n,m})^+,
	\qquad s_{n,m}>0,
\end{equation*}
with a common linear-growth bound.\footnote{
	For example, take the piecewise-linear
interpolant of $e_2$ over a uniform grid with spacing $1/n$ on $[-n,n]$. 
Extend it to the left of $-n$ and to the right of $n$ by affine functions
with the same slopes as the asymptotic slopes of $e_2$.
}
Finite first moments
then allow passage to the limit in all expectations.  We can therefore
conclude that the lower-bound result~(\ref{eq:V2}) holds for an
arbitrary convex payoff $e_2$ of no more than linear growth.
 We can also state the following result.
 
 \begin{proposition}
 	\label{prop:supHsupR}
 	Let $\mu\cx\nu$ be probability measures with finite first moment, 
 	$f$ and $g$ payoffs growing no more than linearly, $g$ convex. Then
 	\begin{equation}
 		\sup_H \left\{
 		\E^{\mu|_{H^c}}[f] + \E^{\Sh^{\nu}(\mu|_H)}[g]
 		\right\} = \sup_{R\geq g}\left\{
 		\E^\mu[f\veeop R] + \E^\nu[g-R]
 		\right\},
 	\end{equation}
 	where the supremum in the left-hand side is over Borel sets $H$ while 
 	the supremum in the right-hand side is over convex functions $R$ of no more than linear growth. 
 \end{proposition}

\section{Three or more exercise dates}
\label{sec:N-majorant}
We proceed to Bermudan options with $N>2$ exercises into payoffs $e_i$ growing no more than linearly and convex for $i>1$. Our goal is to prove that Eq.~(\ref{eq:VN}) gives a sharp lower bound.

The strategy mirrors the Canary case with technical modifications. 
\begin{definition}[Convex-majorant bound]
	\label{def:vcvx}
We define the convex-majorant bound as a supremum over tuples $\mathbf{R} =(R_1,\ldots,R_{N-1})$ of convex functions of no more than linear growth
satisfying a sequence of dominating constraints,
\begin{equation}
	\Vcvx  = \sup_{\substack{R_1 \geq e_2 \veeop R_2\\ \vdots \\ R_{N-2} \geq e_{N-1}\veeop R_{N-1}\\ R_{N-1}\geq e_N}} C(R_1,\ldots,R_{N-1}),
	\label{eq:VNcvx}
\end{equation}
where
\begin{equation}
	C(R_1,\ldots,R_{N-1})  =  \E^{\mu_1}[e_1\veeop R_1] + \sum_{i=2}^{N-1} \E^{\mu_{i}}[e_{i}\veeop R_i - R_{i-1}] + \E^{\mu_N}[e_N-R_{N-1}].
	\label{eq:CRi}
\end{equation}
We will call admissible functions $R_i$ convex majorants. 
\end{definition}

We prove first that $\Vcvx \leq V^*$ by constructing an explicit semi-static subhedge. After that, we again seek to exchange the order of the infimum over measures and the supremum over stopping rules and are driven to introduce convexified problems with relaxed stopping rules. We find an auxiliary relaxed problem for which Sion's theorem applies and use it to establish a relaxed upper bound $\Vrlx\geq V^*$.      

Unlike the Canary option case, we are not immediately able to minimize over measures in the relaxed problem and get a counterpart of Eq.~(\ref{eq:supinfV}). However, we can generalize the construction with a recursion over shadow measures and establish another explicit bound $\Vsh\geq \Vrlx$. We are  finally able to adapt the reasoning of Sect.~\ref{sect:gctp} and prove that $\Vsh\leq \Vcvx$, thereby closing the chain of inequalities and proving that $V^*=\Vcvx=\Vsh$. In this way, we reach the main result without relying on the absence of minimax gap for the original problem, and at this point we will be able to obtain the absence of the gap as a corollary.

\subsection{Convex-majorant bound and semi-static subhedge}
\label{sec:subhedgeN}

We restate the hedging argument of Sect.~\ref{sect:canary_bounds} as a lemma.\footnote{
	We could have used this lemma in Sect.~\ref{sect:canary_bounds} but we chose to give a slightly different derivation optimizing over all continuation sets $H$ and not just those of a special form.  
}
\begin{lemma}[Convex subhedge]
	\label{lemma:subhedge}
	Let $Q$ be a martingale coupling of $(X,Y)$, and let $f$, $g$, and
	$R$ be functions of no more than linear growth, with $g$ and $R$
	convex and $R\geq g$.  Set
	$H=\{x:R(x)>f(x)\}$.
	Then 
	\begin{equation}
		\E^Q[\1_{H^c}(X)f(X) + \1_H(X) g(Y)\mid X] \geq (f\veeop R)(X) + \E^Q[g(Y)-R(Y)\mid X].
		\label{eq:lemmaR}
	\end{equation}
\end{lemma} 	
\begin{proof}
	For any $x$ choose a measurable 
	subgradient $\Delta(x)$ for $R$, so that for any $x,y$,
	\begin{equation}
		R(y)-R(x)\geq \Delta(x) (y-x). 
		\label{eq:subgrad}
	\end{equation}
	Then
	\begin{equation}
		\1_{H^c}(X)f(X)+\1_H(X)g(Y) \geq (f\veeop R)(X) + g(Y) -R(Y) +\1_H(X)\Delta(X)(Y-X).
	\end{equation} 
	Indeed, for $X\in H$ the inequality reduces to Eq.~(\ref{eq:subgrad}) while for $X\in H^c$ it is true by assumption $R\geq g$. Taking conditional expectation given $X$ eliminates the term with
	$Y-X$ by the martingale property. The same argument remains valid with conditioning on
any $\sigma$-algebra $\mathcal G$ such that $X$ is
$\mathcal G$-measurable and $\E^Q[Y\mid\mathcal G]=X$.
\end{proof}
\begin{definition}[Stopping rule associated with convex majorants]
	\label{def:tauR}
	We associate a stopping rule $\tau_{\mathbf{R}}$ with a tuple $\mathbf{R}= (R_1,\ldots,R_{N-1})$
	where each $R_i$ is either a Borel function or a symbol $+\infty$. If $+\infty$ entries are present, they must fill a contiguous initial segment $R_1,\ldots,R_f$ (if no $+\infty$ entries are present, we set $f=0$). 
	 On the path $X_1,\ldots,X_N$, the option is exercised into $e_i(X_i)$ corresponding to the first $i$ with $f<i<N$ such that $R_i(X_i)\leq e_i(X_i)$. If no such $i$ exists, the option is exercised into $e_N$.
\end{definition}

\begin{lemma}
	\label{lemma:etauRC}
Let $N-1$ convex functions of no more than linear growth $R_1,\ldots,R_{N-1}$ be such that $R_i\geq e_{i+1}\veeop R_{i+1}$ for $1\leq i < N-1$ and $R_{N-1} \geq e_N$. The following inequality holds under any martingale coupling $Q$ with marginals $\mu_i$,
\begin{equation}
	\E^Q[e_{\tau_{\mathbf{R}}}(X_{\tau_{\mathbf{R}}})] \geq C(R_1,\ldots,R_{N-1}),
	\label{eq:eq_etaur}
\end{equation}
where the right-hand side is defined by Eq.~(\ref{eq:CRi}).
\end{lemma}
\begin{proof}
Let $H_i=\{x:R_i(x)>e_i(x)\}$. 
Then
\begin{align}
	e_{\tau_{\mathbf{R}}}(X_{\tau_{\mathbf{R}}}) &= \1_{H_1^c}(X_1)e_1(X_1) \nonumber \\
	&+ \1_{H_1}(X_1)\1_{H_2^c}(X_2)e_2(X_2) \nonumber \\
	& \quad\vdots \nonumber \\
	&+ \left(\prod_{i=1}^{N-2}\1_{H_i}(X_i)\right)\1_{H_{N-1}^c}(X_{N-1})e_{N-1}(X_{N-1}) \nonumber \\
	&+
	\left(\prod_{i=1}^{N-1}\1_{H_i}(X_i)\right)e_{N}(X_{N}).
	\label{eq:etaur}
\end{align}
We will take the expectation over $Q$ as a sequence of conditional expectations over ${\mathcal F}_i = \sigma(X_1,\ldots,X_i)$ with $i=N-1,N-2,\ldots,1$.

Applying Lemma~\ref{lemma:subhedge} sequentially for averaging the right-hand side of Eq.~(\ref{eq:etaur}) over $X_{N},\ldots,X_2$, we obtain truncated versions of the same expression with an adjustment in the last payoff $e_i \to e_i \veeop R_i$ until reaching the $e_1 \veeop R_1$ and the remaining average over $X_1$. We will also accumulate a tail of terms $\E^Q\left[\left(\prod_{i=1}^{N-2}\1_{H_i}(X_i)\right)(e_N-R_{N-1})\right]$,
$\E^Q\left[\left(\prod_{i=1}^{N-3}\1_{H_i}(X_i)\right)(e_{N-1}\veeop R_{N-1}-R_{N-2})\right]$, etc. All these terms are nonpositive due to the assumption on the choice of the functions $R_i$, therefore dropping the $0$--$1$ indicator weights given by the products of indicators can only further decrease the sum. We end up with Eq.~(\ref{eq:eq_etaur}).
\end{proof}
Taking the supremum over the admissible tuples of convex majorants in Eq.~(\ref{eq:eq_etaur}) and noticing that stopping times $\tau_{\mathbf{R}}$ are a subset of all stopping times leads to
\begin{equation*}
	\sup_\tau V(Q,\tau) \geq \sup_{\mathbf{R}} V(Q,\tau_{\mathbf{R}}) \geq \Vcvx.
\end{equation*}
Taking the infimum over $Q$ yields
$V^*\geq \Vcvx$.
 
\subsection{Relaxed stopping rules and minimax}
\label{sec:relaxed_rulesN}
The next step in our program is the introduction of relaxed stopping rules\footnote{
The remarks about the interpretation of relaxed stopping rules as either randomized stopping rules or fractional liquidation strategies of \cite{BayraktarZhou2016} made in Sect.~\ref{sect:randomized_canary} apply without modification.
} and the corresponding optimization problems. 
\begin{definition}[Relaxed stopping rules and payoffs]
	\label{def:relaxedrules}
A relaxed stopping rule ${\bolda}$ is a tuple of Borel functions $a_i : {\mathbb R}^i \to [0,1]$, $i=1,\ldots,N-1$, such that
\begin{equation}
0\le a_{i}(x_1,\ldots,x_{i-1},x_{i})\le a_{i-1}(x_1,\ldots,x_{i-1})\le1,
\qquad 1<i<N.
\label{eq:survival-system}
\end{equation}
The relaxed payoff associated with the path $X_1,\ldots,X_N$ is
\begin{equation}
G_{\bolda}(X_1,\ldots,X_N)
=
\sum_{i=1}^{N-1}
\bigl(a_{i-1}(\mathbf{X}_{i-1})-a_i(\mathbf{X}_i)\bigr)e_i(X_i)
+
a_{N-1}(\mathbf{X}_{N-1})e_N(X_N),
\label{eq:randomized-payoff}
\end{equation}
where $\mathbf{X}_i$ is the truncated tuple $(X_1,\ldots,X_i)$ for $1\leq i<N$ and 
$a_0(\mathbf{X}_0)\equiv1$. 
\end{definition}

The value of the option under measure $Q$ and relaxed stopping rule $\bolda$ is denoted
\begin{equation}
	\label{eq:vrlx_bolda}
	\vqr(Q,\bolda) = \E^Q[G_{\bolda}],
\end{equation}
and the relaxed optimization problem is 
\begin{equation*}
	\inf_Q \sup_{\bolda} \vqr(Q,\bolda).
\end{equation*}
We limit our attention to two classes of relaxed stopping rules
\begin{itemize}
	\item continuous, $\mathcal C$, with continuous functions $a_i$;
	\item quantized, $\bar{\mathcal C}$, with functions $\bar a_i$ taking values in a common finite set.
\end{itemize}

\begin{definition}[Level-crossing stopping times, quantized case]
	\label{def:level-crossing-tau}
	Let $\bar{\bolda}$ be a quantized relaxed stopping rule taking a finite number of values $0=\lambda_0<\lambda_1<\cdots<\lambda_M=1$.
We define level-crossing stopping times $\tau_k^{\bar{\bolda}}$ for $1\leq k\leq M$.  On a path $X_1,\ldots,X_N$ we have $\tau_k^{\bar{\bolda}}=i$ if $i$ is the first index such that ${\bar a}_i(X_1,\ldots,X_i)<\lambda_k$ while $\tau_k^{\bar{\bolda}}=N$ if there is no such index, that is, $\bar a_{N-1}(X_1,\ldots,X_{N-1})\geq\lambda_k$.
\end{definition}
The relaxed payoff admits a decomposition into payoffs at the stopping times $\tau_k^{\bar{\bolda}}$,
\begin{equation}
	G_{\bar{\bolda}}(X_1,\ldots,X_N)
	=
	\sum_{k=1}^M
	(\lambda_k-\lambda_{k-1})\,e_{\tau_k^{\bar{\bolda}}}(X_{\tau_k^{\bar{\bolda}}}).
	\label{eq:finite-threshold-average}
\end{equation}

\begin{definition}[Level-crossing stopping times, continuous case]
	\label{def:level-crossing_cont_tau}
With a continuous relaxed stopping rule $\bolda$ and $u\in(0,1]$, define 
$\tau_u^{\bolda}$ on the path $(X_1,\ldots,X_N)$ as
\begin{equation}
	\tau_u^{\bolda}
	=
	\inf\left\{
	i\in\{1,\ldots,N-1\}:
	a_i(\mathbf{X}_i)<u
	\right\},
	\label{eq:threshold-stopping-time}
\end{equation}
if $a_{N-1}(X_1,\ldots,X_{N-1})<u$ and $\tau_u^{\bolda}=N$ otherwise.
\end{definition}
Eq.~(\ref{eq:finite-threshold-average}) has a continuous counterpart,
\begin{equation}
	G_{\bolda}(X_1,\ldots,X_N)
	=
	\int_0^1
	e_{\tau_u^{\bolda}}(X_{\tau_u^{\bolda}})\,du.
	\label{eq:relaxed-payoff-mixture}
\end{equation}

After taking expectations, Eq.~(\ref{eq:relaxed-payoff-mixture})
coincides with the classical mixture representation of a randomized
stopping time; cf.~\cite[Eq.~(1.4)]{EdgarMilletSucheston1982}.
Eq.~(\ref{eq:finite-threshold-average}) is its finite-valued
specialization.

\begin{proposition}[Relaxation of stopping rules]
	\label{prop:fixed-model-relaxation}
Either kind of relaxation does not change the optimal value under a particular measure
\begin{equation}
	\sup_{\bolda\in {\mathcal C}}\vqr(Q,\bolda) = \sup_{\bar{\bolda}\in \bar{\mathcal C}}\vqr(Q,\bar{\bolda})= \sup_\tau V(Q,\tau).
	\label{eq:supa_eq_suptau}
\end{equation}
\end{proposition}
\begin{proof}
By Eqs.~(\ref{eq:finite-threshold-average}), (\ref{eq:relaxed-payoff-mixture}) the payoff under any relaxed rule is a mixture of ordinary stopping rules. Also, any ordinary stopping rule can be regarded as a quantized relaxed stopping rule and can be approximated by continuous relaxed stopping rules, so no gap is possible between the values in Eq.~(\ref{eq:supa_eq_suptau}). 
\end{proof}

The advantage of the continuous class is the absence of the minimax gap for the relaxed problem by Sion's minimax theorem,
\begin{equation}
	\sup_{\bolda\in {\mathcal C}}  \inf_Q \vqr(Q,\bolda) = \inf_Q \sup_{\bolda\in {\mathcal C}} \vqr(Q,\bolda).
	\label{eq:minimaxN}
\end{equation}
 Indeed, the space of admissible measures $Q$ is convex and compact in the $1$-Wasserstein metric, the set of admissible relaxed rules $\bolda$ is convex, the functional $\vqr(Q,\bolda)$ is affine in both arguments and continuous.
 In view of Eqs.~(\ref{eq:supa_eq_suptau}--\ref{eq:minimaxN}) (cf.~\cite[Remark~3.3]{BayraktarZhou2016}),
 \begin{equation}
 	V^* = \inf_Q\sup_\tau V(Q,\tau) = \sup_{\bolda\in {\mathcal C}} \inf_Q \vqr(Q,\bolda).
 	\label{eq:supaCinfQ}
 \end{equation}
 
The step taken in the next section, however, is technically simpler under quantized relaxed stopping rules. In contrast with the case of two exercises of Sect.~\ref{sect:randomized_canary}, here we are not directly able to apply Sion's theorem with discontinuous exercise rules,\footnote{Unlike the case of two exercises, there is no space in which to approximate $\bolda$ by a continuous function uniformly over $Q$.} and we need an additional argument to pass from a supremum over continuous rules to that over quantized rules in the right-hand side of Eq.~(\ref{eq:supaCinfQ}). This requires an approximation that is uniform in $Q$ and does not follow from Eq.~(\ref{eq:supa_eq_suptau}). 

\begin{proposition}[Uniform quantization of relaxed stopping rules]
	\label{prop:continuous-to-quantized}
	\begin{equation}
		\sup_{\bolda\in {\mathcal C}}\inf_Q \vqr(Q,\bolda) \leq \sup_{\bar{\bolda} \in \bar{\mathcal C}}\inf_Q \vqr(Q,\bar{\bolda}).
		\label{eq:a_abar}
	\end{equation}
\end{proposition} 
\begin{proof}
A continuous exercise rule $\bolda$ can be uniformly approximated by a quantized rule $\bar{\bolda}$  taking a finite number of values $0=\lambda_0<\lambda_1<\cdots<\lambda_M=1$. With $\bar\lambda = \max_{1\le i\le M}(\lambda_i-\lambda_{i-1})$, we can ensure that $|a_i-\bar a_i|\leq \bar\lambda$. The relaxed payoff under $\bolda$ is then approximated uniformly in $Q$,
\begin{equation}
	|\vqr(Q,\bolda) - \vqr(Q,\bar{\bolda})| \leq \bar\lambda  	\sum_{i=1}^{N-1}
	\left(
	\E^{\mu_i}[|e_i|]
	+
	\E^{\mu_{i+1}}[|e_{i+1}|]
	\right).
	\label{eq:a_a_bar_approx}
\end{equation}
The right-hand side of Eq.~(\ref{eq:a_a_bar_approx}) tends to 0 as $\bar\lambda \to 0$, so we obtain Eq.~(\ref{eq:a_abar}).
\end{proof}

We take the right-hand side of Eq.~(\ref{eq:a_abar}) as the definition of the relaxed upper bound $\Vrlx$. So far, we proved $V^*\leq \Vrlx$; we will eventually prove that $V^*=\Vrlx$, from which it would follow that an equality holds in Eq.~(\ref{eq:a_abar}).
\subsection{Backward induction on submeasures}
\label{sec:recursion-first}

Motivated by the Canary option case, we expect that shadow measures will be useful in evaluating $\inf_Q \vqr(Q,\bar{\bolda})$ for more than two exercises, leading to a counterpart of Eq.~(\ref{eq:supinfV}). One complication is that we should expect to encounter more general shadows than those of the kind $\Sh^{\beta}(\alpha|_H)$. This is because the exercise decision at $i>1$ is based on the entire path $X_1,\ldots,X_i$ and not only on the last value, so we will generally have continuation submeasures instead of continuation sets. This motivates generalizing the one-step construction of Eq.~(\ref{eq:supinfV}),
\begin{equation*}
	\sup_H\left\{ \E^{\mu_1|_{H^c}}[e_1] + \E^{\Sh^{\mu_2}(\mu_1|_H)}[e_2]\right\},
\end{equation*}
to
\begin{equation*}
	\sup_{0\leq\xi\leq \mu_1}\left\{ \E^{\mu_1-\xi}[e_1] + \E^{\Sh^{\mu_2}(\xi)}[e_2]
	\right\}.
\end{equation*}   
This extends the optimization to arbitrary submeasures, rather than
only those of the form $\xi=\mu_1|_H$.

\begin{definition}[Backward shadow bound]
	\label{def:vshadow}
For a finite measure $\mu\leq \mu_N$ we define
\begin{equation}
W_N(\mu)=\E^{\mu}[e_N],
\label{eq:WN}
\end{equation}
then recursively for $i<N$, $\mu\leq\mu_i$,
\begin{equation}
W_i(\mu)
=
\sup_{0\le\xi\le\mu}
\left\{
\E^{\mu-\xi}[e_i]
+
W_{i+1}\bigl(\Sh^{\mu_{i+1}}(\xi)\bigr)
\right\}.
\label{eq:shadow-recursion}
\end{equation}
The recursion ends with $W_1(\mu)$, and we call $\Vsh=W_1(\mu_1)$ the backward shadow bound. 
\end{definition}

By definition, $\Vsh$ depends only on the marginals $\mu_i$ and payoffs $e_i$ but not on
any specific coupling or stopping rule, so it is a model-free bound.
We will prove that $\Vrlx \leq \Vsh$ in the rest of this section and  $\Vsh=\Vcvx$ in the next section.

We will now construct a particular martingale coupling $Q^{\bar{\bolda}}$ associated with a given quantized relaxed stopping rule\footnote{
	We will see that $Q^{\bar{\bolda}}$ is adverse to the relaxed stopping rule $\bar{\bolda}$ in the sense that $\vqr(Q^{\bar{\bolda}},\bar{\bolda})$ is bounded above by $\Vsh$ independent of $\bar{\bolda}$.
} ${\bar{\bolda}}$ taking values $0=\lambda_0<\lambda_1<\cdots<\lambda_M=1$ and a set of marginals $\mu_i$. (The payoffs $e_i$ do not participate in this construction.) We start with $\mu_1$. The transition kernels to later marginals are defined by forward induction.
 At each exercise date $T_i$ before the last, the marginal $\mu_i$ is decomposed into measures $\xi_{i,k}$ corresponding to each of the possible values $\lambda_k$ of the exercise rule component ${\bar a}_i(X_1,\ldots,X_{i})$. The transition to the states corresponding to the marginal $\mu_{i+1}$ takes each measure $\beta_{i,k} = \sum_{j=k}^M\xi_{i,j}$ to its shadow $\Sh^{\mu_{i+1}}(\beta_{i,k})$.
 
 This is an adaptation of the multiperiod
 shadow-kernel construction of Nutz et al.
 \cite{NutzStebeggTan2020}, where ordered pieces
 of a source marginal are transported successively to their shadows in
 the remaining target marginal, the resulting martingale transports are
 disintegrated into kernels, and the kernels are composed through the
 successive prescribed marginals.
 In our version, the ordering is not fixed in advance but is generated 
 dynamically by the 
 levels of the quantized stopping rule.  

We begin with the distribution $P_1=\mu_1$ at $T_1$ and the exercise rule component ${\bar a}_1(X_1)$. Let 
\begin{equation}
\xi_{1,k} = P_1|_{\left\{ {\bar a}_1(x_1)=\lambda_k\right\}}.
\label{eq:xi_1k}
\end{equation} 

We process the layers from the highest survival level downwards, setting
\begin{align*}
\eta_{1,M} &= \Sh^{\mu_2}(\xi_{1,M}),\nonumber \\
\eta_{1,M-1} &= \Sh^{\mu_2-\eta_{1,M}}(\xi_{1,M-1}),\nonumber \\
&\vdots \nonumber \\
\eta_{1,0} &= 	\Sh^{\mu_2-\sum_{j=1}^M\eta_{1,j}}(\xi_{1,0}).\nonumber \\
\end{align*}	
	Repeated shadow associativity gives, for every $k=0,\ldots,M$,
\begin{equation*}
	\sum_{j=k}^M\eta_{1,j}
	=
	\Sh^{\mu_{2}}\!\left(\sum_{j=k}^M\xi_{1,j}\right).
\end{equation*}
At $k=0$, the source on the right is $\mu_1$; since $\mu_1\cx\mu_{2}$ and the two measures have equal mass, its shadow is all of $\mu_{2}$. Hence
\begin{equation*}
\sum_{j=0}^M\eta_{1,j}=\mu_{2}.
\end{equation*}

For every layer $k$ with nonzero mass of $\xi_{1,k}$, we choose a martingale coupling $\pi_{1,k}$
and disintegrate it as
\begin{equation}
\pi_{1,k}(dx_1,dx_2)
=
\xi_{1,k}(dx_1)\zeta_{1,k}(x_1,dx_2).
\end{equation}
(On zero-mass layers $\zeta_{1,k}$ can be chosen arbitrarily.) We define the full transition kernel $K_1(x_1,dx_2)$ by gluing the layers,
\begin{equation*}
	K_1(x_1,dx_2)
	=
	\zeta_{1,k}(x_1,dx_2)
	\quad\text{when}\quad {\bar a}_1(x_1)=\lambda_k.
\end{equation*}
This gives a martingale coupling $P_2$ on $(X_1,X_2)$,
\begin{equation*}
	P_2(dx_1, dx_2) = P_1(dx_1) K_1(x_1,dx_2).
\end{equation*}	
 
The construction proceeds inductively. If the martingale coupling $P_{i}$ on $(X_1,\ldots,X_{i})$  has already been constructed for $i<N$, we split the marginal $\mu_i$ into layers $\xi_{i,k}$ defined on Borel sets $B$ by
\begin{equation}
	\xi_{i,k}(B) = P_i\left\{(x_1,\ldots,x_i) : {\bar a}_i(x_1,\ldots,x_i)=\lambda_k, x_i \in B\right\}.
	\label{eq:xi_ik}
\end{equation}
For each layer, we produce the shadow in $\mu_{i+1}$ going downwards,
\begin{align*}
	\eta_{i,M} &= \Sh^{\mu_{i+1}}(\xi_{i,M}),\nonumber \\
	\eta_{i,M-1} &= \Sh^{\mu_{i+1}-\eta_{i,M}}(\xi_{i,M-1}),\nonumber \\
	&\vdots \nonumber \\
	\eta_{i,0} &= 	\Sh^{\mu_{i+1}-\sum_{j=1}^M\eta_{i,j}}(\xi_{i,0}),\nonumber \\
\end{align*}	
and we have 
\begin{equation*}
	\sum_{j=k}^M\eta_{i,j}
	=
	\Sh^{\mu_{i+1}}\!\left(\sum_{j=k}^M\xi_{i,j}\right),
\end{equation*}
due to shadow associativity. We disintegrate the martingale coupling on all layers with nonzero mass,
\begin{equation*}
	\pi_{i,k}(dx_i,dx_{i+1})
	=
	\xi_{i,k}(dx_i)\zeta_{i,k}(x_i,dx_{i+1}),
\end{equation*}
and glue the layers into a complete kernel $K_{i}$,
\begin{equation*}
	K_{i}(x_1,\ldots,x_i,dx_{i+1}) = \zeta_{i,k}(x_i,dx_{i+1}) 
	\quad\text{when}\quad {\bar a}_i(x_1,\ldots,x_i)=\lambda_k.
\end{equation*}
 Note that the kernel can depend on the full history and not merely on $x_i$ 
 because we allow history-dependent relaxed stopping rules.
This leads to a martingale coupling $P_{i+1}$ on $(X_1,\ldots,X_i,X_{i+1})$,
 \begin{equation*}
 	P_{i+1}(dx_1,\ldots,dx_i, dx_{i+1}) = P_i(dx_1,\ldots,dx_i) K_{i}(x_1,\ldots,x_i,dx_{i+1}).
 \end{equation*}	
The last step of the induction produces the full martingale coupling $Q^{\bar{\bolda}}=P_N$. 

The next proposition establishes that $Q^{\bar{\bolda}}$ realizes an adverse coupling for a relaxed stopping rule $\bar{\bolda}$ and explains the purpose of defining the backward shadow bound $\Vsh$.
\begin{proposition}
	\label{prop:qaavsh}
\begin{equation}
	\vqr(Q^{\bar{\bolda}},\bar{\bolda}) \leq \Vsh.
	\label{eq:qbar_gbar_leqw}
\end{equation}
\end{proposition}

\begin{proof}
We will prove that the inequality 
\begin{equation*}
	\E^{Q^{\bar{\bolda}}}[e_{\tau_k^{\bar{\bolda}}}(X_{\tau_k^{\bar{\bolda}}})] \leq \Vsh,
\end{equation*}
holds for each individual level-crossing stopping time $\tau_k^{\bar{\bolda}}$ (Definition~\ref{def:level-crossing-tau}). This will prove the assertion~(\ref{eq:qbar_gbar_leqw}) since the weights in Eq.~(\ref{eq:finite-threshold-average}) sum up to 1.

We define the following submeasures of $\mu_i$, 
\begin{align*}
	\alpha_{i,k}(B)
	&=
	Q^{\bar{\bolda}}\left\{(x_1,\ldots,x_N) : x_i\in B,\ \tau_k^{\bar{\bolda}}\ge i\right\},  \\
	\beta_{i,k}(B)
	&=
	Q^{\bar{\bolda}}\left\{(x_1,\ldots,x_N) : x_i\in B,\ \tau_k^{\bar{\bolda}}> i\right\}.
\end{align*}
Comparison with the definitions~(\ref{eq:xi_1k}), (\ref{eq:xi_ik}) of the measures $\xi_{i,k}$ introduced during the construction of $Q^{\bar{\bolda}}$ shows that 
\begin{equation*}
	\beta_{i,k}=\sum_{j=k}^M \xi_{i,j}.
\end{equation*}
Furthermore, $Q^{\bar{\bolda}}$ was constructed from transition kernels in such a way that 
\begin{equation}
	\alpha_{i+1,k}
	=
	\sum_{j=k}^M\eta_{i,j}
	=
	\Sh^{\mu_{i+1}}(\beta_{i,k}).
	\label{eq:finite-alpha-next-shadow}
\end{equation}

Intuitively, $\alpha_{i,k}$ is the measure on $X_i$ for all paths that reached exercise $T_i$
and possibly exercised there, while $\beta_{i,k}$ is the measure on $X_i$ of the paths that progressed beyond $T_i$ without exercise.  

We introduce the tail of the payoff starting from $T_i$,
	\begin{equation*}
		J_{i,k}
		=
		\E^{Q^{\bar{\bolda}}}\!\left[
		\mathbf{1}_{\{\tau_k^{\bar{\bolda}}\ge i\}}
		e_{\tau_k^{\bar{\bolda}}}(X_{\tau_k^{\bar{\bolda}}})
		\right].
	\end{equation*}
	At maturity (see Eq.~(\ref{eq:WN}) for definition of $W_N$),
	\begin{equation*}
	J_{N,k}=\E^{\alpha_{N,k}}[e_N]=W_N(\alpha_{N,k}).
	\end{equation*}
	Going backwards for $i<N$,
	\begin{equation*}
		J_{i,k}
		=
		\E^{\alpha_{i,k}-\beta_{i,k}}[e_i]
		+
		J_{i+1,k}.
	\end{equation*}
	The definition~(\ref{eq:shadow-recursion}) of $W_i$ and Eq.~(\ref{eq:finite-alpha-next-shadow})
	prove the backward induction step from an assumption $J_{i+1,k}\le W_{i+1}(\alpha_{i+1,k})$,
	\begin{align*}
		J_{i,k}
		&\le
		\E^{\alpha_{i,k}-\beta_{i,k}}[e_i]
		+
		W_{i+1}\bigl(\alpha_{i+1,k}\bigr)
		\nonumber \\
			&=
		\E^{\alpha_{i,k}-\beta_{i,k}}[e_i]
		+
		W_{i+1}\bigl(\Sh^{\mu_{i+1}}(\beta_{i,k})\bigr) \nonumber \\
		&\le 
		\sup_{0\leq\xi\leq \alpha_{i,k}}\left(
\E^{\alpha_{i,k}-\xi}[e_i] + 
	W_{i+1}\bigl(\Sh^{\mu_{i+1}}(\xi)\bigr)\right) \nonumber \\
		&= W_i(\alpha_{i,k}).
	\end{align*}
	Upon reaching $i=1$, the induction gives the required inequality,
	\begin{equation*}
	\E^{Q^{\bar{\bolda}}}[e_{\tau_k^{\bar{\bolda}}}(X_{\tau_k^{\bar{\bolda}}})]
	=J_{1,k}
	\le W_1(\mu_1),
	\end{equation*}
completing the proof of Eq.~(\ref{eq:qbar_gbar_leqw}).
\end{proof}

An inequality between $\Vrlx$ and $\Vsh$ immediately follows from Proposition~\ref{prop:qaavsh},
\begin{equation}
	\Vrlx = \sup_{\bar{\bolda} \in \bar{\mathcal C}} \inf_Q \vqr(Q,\bar{\bolda}) \leq 
	\sup_{\bar{\bolda} \in \bar{\mathcal C}} \vqr(Q^{\bar{\bolda}},\bar{\bolda})\leq \Vsh.
	\label{eq:vrlx_leqw}
\end{equation}

\begin{remark}[Continuous shadow lift] 
	\label{remark:cont_lift}
	We could stay with continuous relaxed stopping rules $\bolda$ and prove the existence of $Q^{\bolda}$ such that $\vqr(Q^{\bolda},\bolda)\leq \Vsh$ using Theorem 1.1 from \cite{BeiglbockJuillet2021} to justify martingale shadow lift with continuous thresholds.  We chose to present a more elementary construction with quantized relaxed stopping rules that only requires shadow associativity, which is sufficient by virtue of Proposition~\ref{prop:continuous-to-quantized}.
\end{remark}

\subsection{Identification with convex-majorant optimization}
\label{sec:W-to-majorants}
We generalize the discussion of
Sects.~\ref{sec:two-date-affine} and~\ref{sect:gctp}, leading to
Proposition~\ref{prop:supHsupR}, from submeasures given by restrictions
to subsets to arbitrary submeasures.
 
	\begin{lemma}[Shadow by convex optimization]
		\label{lemma:shsupR}
		Let $\nu_0\cx\nu$ be probability measures with finite first moment, $\mu$ a submeasure of $\nu_0$,
		$g$ convex and growing no more than linearly. Then
		\begin{equation}
			\E^{\Sh^{\nu}(\mu)}[g] = 
			\sup_{R\geq g}\left\{
			\E^\mu[R] + 
			\E^\nu[g-R]
			\right\},
			\label{eq:shsupR}
		\end{equation}
		where the supremum in the right-hand side is over convex functions $R$ of no more than linear growth. 
	\end{lemma}
\begin{proof}
	By definition of the shadow measure, $\mu\cx \Sh^{\nu}(\mu)$, therefore $\E^\mu[R]\leq \E^{\Sh^{\nu}(\mu)}[R]$ as $R$ is convex. Also by definition, $\Sh^{\nu}(\mu) \leq \nu$, therefore
	$\E^\nu[g-R]\leq \E^{\Sh^{\nu}(\mu)}[g-R]$ as $g-R\leq 0$. Adding these inequalities and taking the supremum over $R$ gives 
		\begin{equation*}
		\E^{\Sh^{\nu}(\mu)}[g] \geq
		\sup_{R\geq g}\left\{
		\E^\mu[R] + 
		\E^\nu[g-R]
		\right\}.
	\end{equation*}
To get the opposite inequality, we proceed to transform the 
$\E^{\Sh^{\nu}(\mu)}[g]$ in the same way as we did in Sect.~\ref{sect:gctp}. Considering first 
a multikink payoff $g(x)=\sum_m s_m(x-k_m)^+$, we get
\begin{equation}
	\E^{\Sh^{\nu}(\mu)}[g] = \sup_{\substack{l_m\le k_m\le r_m \\ l_m < r_m}}\left\{\E^{\mu}[R_{\mathbf{l},\mathbf{r}}] + \E^{\nu}[g - R_{\mathbf{l},\mathbf{r}}]\right\},
	\label{eq:g_sh_rl_vec}
\end{equation}
with convex functions $R_{\mathbf{l},\mathbf{r}}\geq g$ given by Eq.~(\ref{eq:bridge-majorant}). Then we approximate an arbitrary convex $g$ by a sum of affine and multikink functions and extend the supremum to all convex functions $R\geq g$.
Enlarging that family to all convex majorants can only increase the right-hand side, while the reverse inequality was proved above. Hence equality follows.
\end{proof}

\begin{lemma}[One-step backward induction on submeasures]
	\label{lemma:supXisupR}
Under the conditions of Lemma~\ref{lemma:shsupR}, assuming additionally that $f$ is $\mu$-integrable,
\begin{equation}
	\sup_{0\leq\xi\leq\mu} \left(
\E^{\mu-\xi}[f] + \E^{\Sh^{\nu}(\xi)}[g]
\right) = \sup_{R\geq g}\left(
\E^\mu[f\veeop R] + \E^\nu[g-R]
\right).
\label{eq:supXisupR}
\end{equation}
\end{lemma}
\begin{proof}
	The difference from Proposition~\ref{prop:supHsupR} is that the measure corresponding to the first marginal is truncated from a probability measure to a submeasure  $\mu$ and that the optimization in the left-hand side goes over generic submeasures and not just over restrictions $\mu|_H$. 
	We could again go through the steps of Sect.~\ref{sect:gctp} but we can also substitute the expression~(\ref{eq:shsupR}) for the expectation over shadow measure into 
	the left-hand side of Eq.~(\ref{eq:supXisupR}), interchange the order of suprema over $\xi$ and $R$, and evaluate the inner supremum over measures $\xi$, 
\begin{equation*}
\sup_{0\le\xi\le\mu}
\left\{
\E^{\mu-\xi}[f]+\E^\xi[R]
\right\}
=
\E^\mu[f\vee R].
\end{equation*}
(Indeed, writing $\xi=q\mu$ reduces the optimization to choosing $0\le q(x)\le1$ pointwise according to the sign of $R(x)-f(x)$). 
\end{proof}

We apply Eq.~(\ref{eq:supXisupR}) of Lemma~\ref{lemma:supXisupR}
at each step of the backward recursion~(\ref{eq:shadow-recursion}).  Suppose inductively that $W_{i+1}$ has the majorant
		representation with first continuation payoff
		$e_{i+1}\vee R_{i+1}$.  For fixed later majorants, apply the
		one-step identity with
		\begin{equation*}
		f=e_i,
		\qquad
		g=e_{i+1}\vee R_{i+1}.
		\end{equation*}
		Taking the supremum over the remaining majorants yields the constraint
		$R_i\geq e_{i+1}\vee R_{i+1}$, and we ultimately get
\begin{equation}
\Vsh
=
\sup_{\substack{R_1 \geq e_2 \veeop R_2\\ \vdots \\ R_{N-2} \geq e_{N-1}\veeop R_{N-1}\\ R_{N-1}\geq e_N}} \E^{\mu_1}[e_1\veeop R_1] + \sum_{i=2}^{N-1} \E^{\mu_{i}}[e_{i}\veeop R_i - R_{i-1}] + \E^{\mu_N}[e_N-R_{N-1}],
	\label{eq:W}
\end{equation}
an equation whose right-hand side is the convex-majorant objective
$C(R_1,\ldots,R_{N-1})$ from Definition~\ref{def:vcvx}. Therefore $\Vsh=\Vcvx$.

We have completed the chain $\Vcvx\leq V^*\leq \Vrlx\leq \Vsh = \Vcvx$ and now can claim that all inequalities in it are actually equalities.
We can summarize the results as a theorem
\begin{theorem}[Constructive representations of the model-free lower bound]
	\label{theorem:constructive}
	For any set of $N$ one-dimensional marginals with finite first moment in convex order $\mu_1\cx\cdots\cx\mu_N$ and $N$ payoffs $e_i(x_i)$ of no more than linear growth, convex for $i>1$, the following quantities are equal
	\begin{enumerate}
		\item The model-free lower bound of the Bermudan option $V^*=\inf_Q \sup_\tau \E^Q[e_\tau(X_\tau)]$ with $Q$ ranging over all martingale couplings of the marginals and
		$\tau$ ranging over all stopping times.
		\item The convex-majorant bound $\Vcvx$ of Definition~\ref{def:vcvx}.
		\item The backward shadow bound $\Vsh$ of Definition~\ref{def:vshadow}.
	\end{enumerate}
\end{theorem}

We recapitulate the steps of the proof.
\begin{itemize}
	\item Step 1. $\Vcvx \leq V^*$ by a semi-static hedging argument (Sect.~\ref{sec:subhedgeN}).
	\item Step 2. $V^* = \inf_Q \sup_{\bolda} \vqr(Q,\bolda)$, with continuous relaxed stopping rules $\bolda$ and relaxed values $\vqr(Q,\bolda)$  (Sect.~\ref{sec:relaxed_rulesN}).
	\item Step 3. $\inf_Q \sup_{\bolda} \vqr(Q,\bolda) = \sup_{\bolda} \inf_Q  \vqr(Q,\bolda)$ by Sion's theorem (Sect.~\ref{sec:relaxed_rulesN}).
	\item Step 4. $\sup_{\bolda} \inf_Q  \vqr(Q,\bolda)\leq \sup_{\bar{\bolda}} \inf_Q  \vqr(Q,\bar{\bolda})$, with quantized relaxed stopping rules $\bar{\bolda}$, by an approximation uniform in $Q$ (Eq.~(\ref{eq:a_a_bar_approx})).
	\item Step 5. $\sup_{\bar{\bolda}} \inf_Q  \vqr(Q,\bar{\bolda}) \leq \Vsh$ (Sect.~\ref{sec:recursion-first}).
	\item Step 6. $\Vsh= \Vcvx$ (Sect.~\ref{sec:W-to-majorants}), hence all inequalities in the previous steps hold as equalities.
\end{itemize}
 At this point we can also deduce the absence of a minimax gap.
 \begin{corollary}[$Q$-$\tau$ minimax duality]
 	\label{corollary:duality}
 	\begin{equation}
 			\sup_\tau\inf_Q V(Q,\tau) = \inf_Q \sup_\tau V(Q,\tau).
 	\end{equation}
 \end{corollary}
 	
\begin{proof} 
Taking the infimum over $Q$ in inequality~(\ref{eq:eq_etaur}) first and the supremum over admissible convex majorants next we get
\begin{equation}
	\sup_{\mathbf{R}}\inf_Q V(Q,\tau_{\mathbf{R}}) \geq \sup_{\mathbf{R}}C(R_1,\ldots,R_{N-1}) = \Vcvx.
\end{equation}
Expanding from the set of $\tau_{\mathbf{R}}$ to all stopping times can only increase the supremum,
therefore
\begin{equation}
	\sup_\tau \inf_Q V(Q,\tau) \geq \Vcvx.
	\label{eq:supinfvcvx}
\end{equation}
We already proved that $\Vcvx=V^*=\inf_Q\sup_\tau V(Q,\tau)$ and we have the elementary minimax inequality,
\begin{equation}
	\sup_\tau \inf_Q V(Q,\tau) \leq \inf_Q \sup_\tau V(Q,\tau).
	\label{eq:elemminimax}
\end{equation}
Comparing Eqs.~(\ref{eq:supinfvcvx}) and (\ref{eq:elemminimax}) we obtain the minimax duality.
\end{proof}

Having proved the sharpness of the convex-majorant bound~(\ref{eq:VNcvx}) in the general case, we may ask whether a smaller class of convex majorants could be sufficient to achieve the bound under additional assumptions. In particular, a valid question is whether Piterbarg's affine form can be generalized from the Canary option to more than two exercises into single-kink convex payoffs, such as calls and puts. More specifically, we can ask if we can restrict the optimization in Eq.~(\ref{eq:VNcvx}) to convex functions of the form $R_{N-1}=e_N\veeop \ell_{N}$, $R_{N-2} = e_{N-1}\veeop R_{N-1}\veeop \ell_{N-1}$, $\ldots$, $R_1=e_2\veeop R_2\veeop \ell_2$ with $N-1$ affine functions $\ell_2$, \ldots, $\ell_{N}$. From geometrical intuition gained in Sect.~\ref{sect:gctp}, the answer to this question should be negative because an affine line $\ell_N$ may split the single kink of the payoff $e_N$ into two kinks, so we may need as many as three lines to handle these two kinks and an additional kink of $e_{N-1}$ at the previous exercise date $T_{N-1}$, then as many as  $2\times 3+1$ lines at the $T_{N-2}$, etc. A more general class of piecewise-linear convex majorants~(\ref{eq:bridge-majorant}) will suffice, but the total number of affine segments in the functions of this class may grow exponentially with the number of exercises.

The following reduction of the space of convex majorants will be useful in proving the attainment of an optimal stopping time in Sect.~\ref{sec:majorant-attainment}.

\begin{lemma}[Reduction to Lipschitz convex majorants]
	\label{lemma:rnormalization}
	Let $L$ be the largest absolute value of the asymptotic slopes of all payoff functions
	$e_2,\ldots,e_N$,
	\begin{equation*}
		L
		=
		\max_{1<i\leq N}
		\max\left(
		\left|\lim_{x\to-\infty}\frac{e_i(x)}{x}\right|,
		\left|\lim_{x\to+\infty}\frac{e_i(x)}{x}\right|
		\right).
	\end{equation*}
	The asymptotic slopes exist and are finite because the payoffs are convex
	and of no more than linear growth by assumption.
	The supremum in Eq.~(\ref{eq:VNcvx}) can be restricted to convex
	functions that are Lipschitz with the common constant $L$.
\end{lemma}

\begin{proof}
	We start with an arbitrary admissible tuple
	$(R_1,\ldots,R_{N-1})$ and normalize its components starting from
	$R_{N-1}$ and proceeding backwards.  We remove unnecessary slack
	in each optimization layer of the form
	\begin{equation*}
		\sup_{R\geq g}
		\left(
		\E^{\mu}[f\veeop R]
		+
		\E^\nu[g-R]
		\right),
	\end{equation*}
	with convex $g$, $R$ and $\mu\cx\nu$, following Proposition~2.12
	of \cite{HobsonNorgilas2026}.
	By Lemma~2.3 in \cite{BeiglbockHobsonNorgilas2022},
	$\widehat R=R-(R-g)^c$ is convex, and also
	$\widehat R\geq g$.  The replacement
	$R\mapsto\widehat R$ can only increase the objective because we lose
	in the first term at most $\E^\mu[(R-g)^c]$, while we gain
	$\E^\nu[(R-g)^c]$ in the second term, and the loss is no greater
	than the gain due to the convex order $\mu\cx\nu$.
	Since $\widehat R\leq R$, this replacement also preserves the
	constraint involving the preceding majorant.
	
	If $p\leq\widehat R-g$ is convex, then $p+(R-g)^c$ is a convex minorant of $R-g$. By maximality of $(R-g)^c$, $p\leq0$; since zero is itself a convex minorant, $(\widehat R-g)^c=0$.
	So, after replacing $R$ by
	$\widehat R=R-(R-g)^c$,
	we may restrict the optimization to normalized majorants satisfying
	$(\widehat R-g)^c=0$.

	Every normalized majorant has the same asymptotic slopes at
	$\pm\infty$ as $g$; otherwise a nonzero one-sided affine hinge
	would be a convex minorant of $\widehat R-g$.
	We start by applying the normalization to $R_{N-1}$ with $g=e_N$
	and proceed backwards, using
	$g=e_{i+1}\veeop\widehat R_{i+1}$
	when normalizing $R_i$.  The absolute values of
	the asymptotic slopes of all normalized majorants remain bounded by
	those of the payoff functions and hence by $L$.
	The proof is completed by noticing that a bound on the absolute
	values of both asymptotic slopes of a convex function also serves
	as a global Lipschitz constant.
\end{proof}

\section{Attainment of minimizing measure and maximizing stopping rules}
\label{sec:majorant-attainment}

We turn to the question of 
attainment of suprema and infima in our minimax problems.
We note first that 
$\sup_\tau V(Q,\tau)$ under a fixed measure $Q$ is always attained, as an optimal stopping rule can be defined by comparing exercise value $e_i$ with continuation value $\E^Q[Y_{i+1}|{\mathcal F}_i]$ in the standard backward induction algorithm ($Y_N=e_N(X_N)$, $Y_i = \max \left\{e_i(X_i),\E^Q[Y_{i+1}|{\mathcal F}_i]\right\}$, $V=\E^Q[Y_1]$).

On the other hand, we cannot claim that $\inf_Q V(Q,\tau)$ is necessarily attained for every fixed stopping rule $\tau$.
This is true from continuity and compactness for the case of two exercises because the stopping rule can be approximated in $L^1(\mu_1)$ uniformly in $Q$
but the argument does not extend to more than two exercises.\footnote{A counterexample for three exercise dates with a stopping rule at $T_2$ dependent on both $X_2$ and $X_1$ can be constructed; its presentation is beyond the scope of this paper.} Nevertheless, after optimization over $\tau$ the infimum over $Q$ in $\inf_Q \sup_\tau V(Q,\tau)$ is attained, as is the supremum in $\sup_\tau \inf_Q V(Q,\tau)$.

\begin{theorem}[Existence of minimizing measure and maximizing stopping rule]
	\label{theorem:attainment}
There exists a saddle pair $(Q^*,\tau^*)$ such that 
\begin{equation}
V(Q^*,\tau)\leq V(Q^*,\tau^*)=V^*\leq V(Q,\tau^*),
\end{equation}
for any admissible measure $Q$ and stopping rule $\tau$. 
\end{theorem}
\begin{proof}
	Eq.~(\ref{eq:supa_eq_suptau}) justifies substitution of the inner supremum over stopping times by a supremum over relaxed continuous stopping rules. The function $Q\to \vqr(Q,\bolda)$ is continuous for any continuous rule $\bolda$, hence $Q\to\sup_{\bolda} \vqr(Q,\bolda)$ is lower semicontinuous and therefore attains its minimum on the compact space of admissible measures $Q$ at some $Q^*$.
 
It remains to show the existence of a stopping time $\tau^*$ realizing the supremum in $\sup_\tau \inf_Q V(Q,\tau)$. We proved that  
\begin{equation}
	V^* =  \inf_Q \sup_\tau V(Q,\tau)
	=
	\sup_{\substack{R_1 \geq e_2 \veeop R_2\\ \vdots \\ R_{N-2} \geq e_{N-1}\veeop R_{N-1}\\ R_{N-1}\geq e_N}}
	C(R_1,\ldots,R_{N-1}),
	\label{eq:VstarF}
\end{equation}
where the equivalent form of Eq.~(\ref{eq:CRi}), obtained using
$e_i\veeop R_i=R_i+(e_i-R_i)^+$, is
\begin{equation}
	C(R_1,\ldots,R_{N-1})
	=
	\E^{\mu_N}[e_N]
	+
	\sum_{i=1}^{N-1}
	\left\{
	\E^{\mu_i}[(e_i-R_i)^+]
	+
	\E^{\mu_i}[R_i]
	-
	\E^{\mu_{i+1}}[R_i]
	\right\}.
	\label{eq:FRs}
\end{equation}
Assuming first that the supremum over admissible  functions $\mathbf{R}=(R_1,\ldots,R_{N-1})$ is attained on $\mathbf{R}^*=(R_1^*,\ldots,R_{N-1}^*)$, so $C(R_1^*,\ldots,R_{N-1}^*)=V^*$, an attaining stopping rule $\tau^*$ is $\tau_{\mathbf{R}^*}$ (see Definition~\ref{def:tauR} and Lemma~\ref{lemma:etauRC}).

It might happen that the supremum over tuples $\mathbf{R}$ is not attained, in which case there is only a maximizing sequence $\mathbf{R}^{(1)},\mathbf{R}^{(2)},\ldots$ with $C(\mathbf{R}^{(j)})\to V^*$. We will show that an attaining rule $\tau^*$ exists in this case too, and it corresponds to a subsequence in which
several initial components $R_1,\ldots,R_f$ of the approximating sequence escape to infinity,\footnote{$R_1\geq R_2\geq\cdots\geq R_{N-1}$ by construction, therefore, when any component $R_i$ escapes to infinity, all $R_j$ with $j<i$ do as well.} while the remaining ones (if any) converge to $R_{f+1}^*,\ldots,R_{N-1}^*$. This corresponds to a well-defined stopping rule $\tau_{\mathbf{R}^*}$ with
$R_1^*=\cdots=R_f^*=+\infty$ with mandatory continuation over $T_1,\ldots,T_f$.
 
We present an argument based on compactification of the space of
admissible convex majorants $\mathbf{R}$.  By Lemma~\ref{lemma:rnormalization}, we can work in a subspace of
majorants which have a common Lipschitz constant $L$. 
We decompose each majorant into a centered component and an intercept,
\begin{equation}
	\label{eq:compactR}
R_i=\bar R_i + c_i,
\qquad
c_i=R_i(0),
\qquad
\bar R_i(0)=0.
\end{equation}
For a fixed value of the intercept $c_i$, the corresponding centered
functions $\bar R_i$ are uniformly bounded on every segment $[-n,n]$, since
$|\bar R_i(x)|\leq L|x|$.
The larger family of centered convex functions satisfying
$\bar R_i(0)=0$ and having Lipschitz constant at most $L$ is
equicontinuous and closed under locally uniform convergence. 
The Arzel\`a--Ascoli theorem on each segment
$[-n,n]$, followed by a diagonal argument, shows that this
family is sequentially compact in the topology\footnote{
This topology can be metrized, for example, by
\begin{equation*}
d_{\mathrm{loc}}(\bar R_1,\bar R_2)
=
\sum_{n=1}^{\infty}
2^{-n}
\left(
\sup_{|x|\leq n}|\bar R_1(x)-\bar R_2(x)|\wedge1
\right).
\end{equation*}
} of locally uniform convergence on
$\mathbb R$.  

We compactify the one-dimensional intercept coordinate by adjoining
its own point $c_i=+\infty$ for every centered function $\bar R_i$.
We show that the objective extends
continuously to this compactification. 
Indeed, in the compactified coordinates~(\ref{eq:compactR}),
the objective takes the form
\begin{equation}
	C
	={}
	\E^{\mu_N}[e_N]
	+
	\sum_{i=1}^{N-1}
	\left\{
	\E^{\mu_i}[(e_i-c_i-\bar R_i)^+]
	+
	\E^{\mu_i}[\bar R_i]
	-
	\E^{\mu_{i+1}}[\bar R_i]
	\right\}.
	\label{eq:Ccompactified}
\end{equation}
Here
\begin{equation}
c_i\in[m_i,+\infty],
\qquad
m_i=\max_{i<j\leq N}e_j(0),
\end{equation}
and the positive-part term is defined to be zero when $c_i=+\infty$.
The common Lipschitz bound and finite first moments imply that
$C$ is continuous on the compactified space.

If the first $f$ intercepts are infinite, then
\begin{equation}
C
=
C^{(f)}(R_{f+1},\ldots,R_{N-1})
+
\sum_{i=1}^{f}
\left(
\E^{\mu_i}[\bar R_i]
-
\E^{\mu_{i+1}}[\bar R_i]
\right)
\leq
C^{(f)}(R_{f+1},\ldots,R_{N-1}),
\label{eq:Cf}
\end{equation}
where $C^{(f)}$ is the convex-majorant objective for the truncated
problem beginning at $T_{f+1}$, and $\E^{\mu_i}[\bar R_i]
-
\E^{\mu_{i+1}}[\bar R_i]\leq0$ follows from $\mu_i\cx\mu_{i+1}$.

The argument is now completed by continuity and compactness.
Indeed, after normalization, every maximizing sequence admits a
subsequence such that, for each component, the centered functions
converge locally uniformly and the corresponding intercepts either
converge to a finite value or tend to $+\infty$.  In the first case,
the component converges locally uniformly to a finite convex
majorant.  In the second case, it diverges locally uniformly to
$+\infty$, because its centered part remains uniformly bounded on
every compact interval.  The nesting constraints imply that the
diverging components form an entire initial segment.

Suppose that the length of the diverging initial segment is $f$.
The remaining finite majorants define an ordinary stopping rule
$\tau^*$, which continues through $T_1,\ldots,T_f$ and thereafter
follows the rule generated by $R_{f+1},\ldots,R_{N-1}$.
By the multistep subhedging inequality~(\ref{eq:eq_etaur}) for the
truncated problem and Eq.~(\ref{eq:Cf}),
\begin{equation}
	V^*
	\leq
	C^{(f)}(R_{f+1},\ldots,R_{N-1})
	\leq
	\inf_Q V(Q,\tau^*)
	\leq
	\sup_\tau\inf_Q V(Q,\tau)
	=
	V^*,
\end{equation}
where the last equality follows from
Corollary~\ref{corollary:duality}.  Hence all inequalities are
equalities, and $\tau^*$ attains $V^*$.
\end{proof}

The optimal stopping rule is accompanied by an optimal semi-static
subhedge.  For a finite optimizing tuple of majorants
$\mathbf{R}^*=(R_1^*,\ldots,R_{N-1}^*)$, the static positions have
payoffs
\begin{equation*}
e_1^*=e_1\veeop R_1^*,
\qquad
e_i^*=e_i\veeop R_i^*-R_{i-1}^*,
\quad 2\leq i<N,
\qquad
e_N^*=e_N-R_{N-1}^*,
\end{equation*}
at the respective exercise dates.  If
$\Delta_i(x)\in\partial R_i^*(x)$ is a measurable subgradient, the
dynamic position taken at $T_i$ consists of
\begin{equation*}
\1_{\{\tau^*>i\}}\Delta_i(X_i)
\end{equation*}
units of the underlying.
The initial cost of this strategy is
$C(\mathbf{R}^*)=V^*$.  Iterating the pathwise inequality in
Lemma~\ref{lemma:subhedge} shows that its total payoff is dominated
pathwise by $e_{\tau^*}(X_{\tau^*})$.  Under the minimizing coupling $Q^*$ the
subhedge is tight, so the two payoffs agree $Q^*$-almost surely.  If
the compactified optimizer has an initial segment of infinite
majorants, the same construction is applied to the finite tail after
the corresponding mandatory-continuation dates.

The existence of a saddle pair $(Q^*,\tau^*)$ has now been established. We note that the maximizing stopping rule $\tau^*$ is always local in the sense that it 
uses the past only to determine whether
exercise has already occurred, while the exercise decision at $T_i$ depends
only on the current state $X_i$.
 We do not have a similar property for the minimizing measure $Q^*$, which does not necessarily correspond to a Markovian process. 

\section{Discussion and further directions}
\label{sec:discussion_directions}

\subsection{Connections to previous work}
\label{sec:litrev}

For conventional model-based valuation of Bermudan options,
see Brigo and Mercurio
\cite{BrigoMercurio2006} and Andersen and Piterbarg
\cite{AndersenPiterbarg2010}.  The present work instead considers the
sharp lower value over all martingale models consistent with prescribed
marginal laws.

Among the foundational references for the formulation of
model-independent pricing as a martingale
optimal transport problem are Beiglb\"ock et al.
\cite{BeiglbockHenryLaborderePenkner2013} in discrete time and
Galichon et al.~\cite{GalichonHenryLabordereTouzi2014} in the
continuous-time setting. The duality theory was subsequently developed
in considerable generality by Beiglb\"ock et al.
\cite{BeiglbockNutzTouzi2017}.  
Pricing--hedging dualities for path-dependent
claims, including formulations with several prescribed marginals, were
developed by Dolinsky and Soner \cite{DolinskySoner2015}.
We refer to Henry-Labord\`ere's
monograph \cite{HenryLabordere2017} for an exposition of martingale
optimal transport in the context of model-free pricing and hedging of
financial derivatives, together with a broad account of the early
literature, presented in a form accessible to practitioners in
quantitative finance.

Model-independent bounds for claims with early exercise have been
studied in a number of works, with much of the theoretical development
concerned with optimal-transport dualities, possible failures of minimax
duality, and conditions under which duality can be recovered.
In particular, Bayraktar et al.
\cite{BayraktarHuangZhou2015} showed that the two optimization orders
in the lower-bound problem need not agree in the presence of static
calibration constraints.  Bayraktar and Zhou
\cite{BayraktarZhou2016} obtained hedging dualities after allowing liquidation
strategies under which fractions of the claim may be exercised on
different dates.  Aksamit et al.
\cite{AksamitDengOblojTan2019} recovered pricing--hedging duality by
passing to suitable enlargements of the probabilistic setup.
Continuous-time pricing–hedging duality results for American options, 
using randomized stopping and measure decoupling, were obtained by Guo and 
Ob\l{}\'oj \cite{GuoObloj2025}.
A different approach is pursued by Belomestny et al.
\cite{BelomestnyHubnerKraetschmerNolte2019}, who give sufficient
measure-theoretic and topological conditions for minimax identities
without imposing time consistency or relaxing exercise to liquidation
strategies.  
A closely related use of randomization as a minimax device appears in
Belomestny and Kr\"atschmer
\cite{BelomestnyKraetschmer2016}.

We take a more constructive approach, providing an explicit solution
for the lower bound both in terms of optimization over convex
majorants and in terms of a backward induction on submeasures.
Relaxed stopping rules enter only as a technical tool in intermediate
steps, and the minimax identity is obtained as a consequence of the
explicit characterization of the bound.  The principal limitation of
our result is the convexity assumption on the exercise payoffs.  This
condition is, however, easy to verify and is satisfied by many standard
option payoffs.  
A direct precursor is Piterbarg
\cite{Piterbarg2026Canary}, who proved that the sharp
lower bound for a Canary call is given by the optimal straddle--butterfly
subhedge for arbitrary two-date marginals in convex order.  The revised version of \cite{Piterbarg2026Canary}
also employs shadow measures and derives the
minimax duality via relaxation of the stopping rule.
The present paper extends
the convex-majorant construction to arbitrary finite exercise
horizons and general convex payoff profiles, identifies it with a
backward shadow recursion on dynamically optimized submeasures, and
proves multidate minimax duality and attainment by an ordinary
stopping rule.
 
Hobson and
Norgilas \cite{HobsonNorgilas2026} provide a close upper-bound analogue
 for a Bermudan
option with two exercise dates and convex payoffs $a$, $b$ (for other work on this problem, see also 
Hobson and Neuberger \cite{HobsonNeuberger2017}, Hobson and
Norgilas \cite{HobsonNorgilas2019}, and Lim \cite{Lim2023}).
  They show that the full
superhedging dual can be reduced to an optimization over a
convex function $\psi$ dominating the terminal payoff $b$:
\begin{equation*}
D
=
\inf_{\substack{\psi\geq b\\ \psi\ \text{convex}}}
\left\{
\E^{\mu_1}[(a-\psi)^+]+\E^{\mu_2}[\psi]
\right\},
\end{equation*}
which is structurally similar to our Eq.~(\ref{eq:V2}).  
Hobson and Norgilas also use the normalization
$\psi\longmapsto\psi-(\psi-b)^c$,
which removes convex slack without increasing the superhedging cost.
This is the upper-bound counterpart of the normalization
$R\mapsto R-(R-g)^c$ used as a part of our attainment proof in Lemma~\ref{lemma:rnormalization}.  

Our proofs rely critically on the shadow-measure construction of
Beiglb\"ock and Juillet
\cite{BeiglbockJuillet2016} and on the shadow potential formula of
Beiglb\"ock et al.~\cite{BeiglbockHobsonNorgilas2022}.
Per Remark~\ref{remark:cont_lift}, the backward recursion of
Sect.~\ref{sec:recursion-first} could alternatively be formulated
directly in terms of continuous relaxed stopping rules using the lifted
shadow coupling theorem of Beiglb\"ock and Juillet
\cite{BeiglbockJuillet2021}.  We instead pass through quantized
stopping rules and finite shadow associativity, which keeps the required
shadow machinery more elementary.

Stability of martingale optimal transport under perturbations of the
prescribed marginals was established by Backhoff-Veraguas and Pammer
\cite{BackhoffVeraguasPammer2022}, while Beiglb\"ock et al.
\cite{BeiglbockJourdainMargheritiPammer2023} developed the
corresponding theory for weak martingale optimal transport.
Jourdain and Pammer \cite{JourdainPammer2024} further extended this
framework, obtaining stability results for applications including robust
pricing of American options and shadow couplings. Under suitable
continuity assumptions, our two-date lower-bound problem fits the weak
martingale transport framework of
\cite{BeiglbockJourdainMargheritiPammer2023}.

The backward recursion of Sect.~\ref{sec:recursion-first} is
related to martingale transport constructions based on shadow measures.
Building on Beiglb\"ock and Juillet
\cite{BeiglbockJuillet2016}, Henry-Labord\`ere and Touzi
\cite{HenryLabordereTouzi2016} gave an explicit construction of the
left-monotone martingale transport.  Nutz et al.
\cite{NutzStebeggTan2020} extended the approach to the multiperiod
setting through obstructed shadows.  In these left-monotone
constructions, the distinguished source submeasures are the initial
segments $\mu|_{(-\infty,a]}$.  Br\"uckerhoff et al.
\cite{BruckerhoffHuesmannJuillet2022} develop a more general
shadow-martingale construction in which parametrized submeasures are
propagated by obstructed shadows through a peacock, including peacocks
indexed by continuous time.  In our recursion, by contrast, the
submeasure to be propagated is not prescribed by a fixed
parametrization but is determined dynamically at each exercise date as
part of the optimal-stopping problem.
In a different continuous-time setting without prescribed marginal
constraints, Nutz and Zhang \cite{NutzZhang2015} prove the existence
of the value and an optimal stopping time for a robust
controller--stopper game, and obtain a saddle point under an
additional compactness condition.

\subsection{Further directions}
\label{sec:directions}

Several natural extensions remain for future exploration.
\subsubsection{Continuous exercise.}
	In a continuous exercise setting, we will have a peacock of one-dimensional marginals, $\mu_t$,
	and time-dependent payoff functions $e_{t}$.
	Approximation by a sequence of increasingly refined exercise grids
	produces an increasing sequence of sharp Bermudan lower bounds,
	but it is not obvious what features of the discretized problem will hold 
	in the continuous limit.
	The convex-majorant bound $\Vcvx$ will likely become impractical, while the bound $\Vsh$
	defined by the backward induction on submeasures looks more promising. It would be interesting to investigate whether an infinitesimal limit
	of the recursion~(\ref{eq:shadow-recursion}) admits an HJB-type
	characterization suitable for model-independent pricing in continuous
	time.

\subsubsection{More general payoffs \texorpdfstring{$e_i$}{ei}.}
	The optimization over measures with the aid of shadow measures is the central ingredient of our approach, limiting us to convex payoffs. Extension to nonconvex payoffs would require different tools. It remains to be seen if the problem with nonconvex payoffs admits any explicit constructions of sharp lower bounds. 
	
	An even more ambitious undertaking would be to include path-dependent payoffs, $e_i=e_i(X_1,\ldots,X_i)$.
	One class that is amenable to treatment with the same methods is
additive path dependence with convex $r_j$, $\widetilde e_i$, of the form
\begin{equation*}
e_i(X_1,\ldots,X_i)
=
\sum_{j=1}^{i-1}r_j(X_j)+\widetilde e_i(X_i).
\end{equation*}

	\subsubsection{Numerical implementation and applications.}
	The reduction of the minimax problem to an optimization over convex
majorants or a backward induction on submeasures offers a substantial
reduction in computational burden, although significant challenges
remain for more than a few exercise dates.  The optimization over
majorants $R_i$ is not sequential and must generally be run
simultaneously over all components.  The backward induction 
on submeasures is sequential in time, but it
does not decompose into independent one-step optimizations: the choice
of the continuation submeasure at each date depends on the entire
future value functional.  Thus sequentiality does not by itself imply that
the backward-shadow formulation is numerically less demanding, and
which of the two reduced representations is preferable may depend on
the problem.  Good low-dimensional approximations may be possible,
with either approach ensuring a valid lower bound after restriction of
the optimization range.  The analysis of
Sect.~\ref{sect:gctp} suggests that carefully keeping track of payoff
kinks may help reduce the gap between the theoretical sharp bound and
what is practically accessible in a numerical implementation.

	With efficient numerical implementations, model-free bounds could
	become a useful tool for model-risk mitigation of Bermudan options and
	callable bonds.  Practical applications would also benefit from taking
	bid--ask spreads into account.  One natural extension would replace
	the exactly prescribed marginals by their upper and lower bounds,
	$\mu_i^b\preceq_{\rm cx}\mu_i^a$, as in 
	\cite{LiangNutzShengTissotDaguette2026}.  The model would then be
	allowed to select marginal laws $\nu_i$ satisfying
	\begin{equation*}
	\mu_i^b\preceq_{\rm cx}\nu_i\preceq_{\rm cx}\mu_i^a,
	\qquad
	\nu_1\preceq_{\rm cx}\cdots\preceq_{\rm cx}\nu_N,
	\end{equation*}
	in addition to selecting a martingale coupling of those marginals.
	It remains to determine whether the convex-majorant and backward-shadow
	representations retain comparably explicit forms under this enlarged
	optimization.

\section{Conclusions}
\label{sec:conclusions}
The main result of this work is Theorem~\ref{theorem:constructive}
 providing two representations of the sharp model-free lower bound for a Bermudan option.  
The convex-majorant bound $\Vcvx$ (Definition~\ref{def:vcvx}) is an optimization over
explicit semi-static subhedges, while the backward shadow bound
$\Vsh$ (Definition~\ref{def:vshadow}) is obtained by backward induction on surviving
submeasures, with each continuation submeasure propagated to its
shadow in the next marginal. 

The problem can be regarded as an adversarial game where one player tries to maximize the value of the 
option by choosing a stopping time $\tau$, while the other player tries to minimize the value
by choosing a martingale coupling $Q$ with prescribed one-dimensional marginals.
We proved the absence of the minimax gap (Corollary~\ref{corollary:duality})
and the existence of a saddle point (Theorem~\ref{theorem:attainment}).
The maximizing stopping rule is accompanied by a semi-static subhedge
of cost $V^*$ whose payoff is dominated pathwise by the exercised
option payoff and is tight under the minimizing coupling $Q^*$.
The most significant limitation of the results is the requirement of convexity of
option payoffs from the second exercise date onward.
No regularity assumptions are imposed on the marginals beyond finite
first moments and convex ordering, which ensures the existence of at
least one martingale coupling.
 
We use the shadow measures \cite{BeiglbockJuillet2016} as the main technical tool and derive
some auxiliary results potentially valuable beyond the Bermudan-option problem,
in particular Lemmas~\ref{lemma:shsupR} and \ref{lemma:supXisupR}. 
The former expresses the expectation of a convex function over a shadow measure
as a supremum over convex functions,
\begin{equation*}
\E^{\Sh^\nu(\mu)}[g]
=
\sup_{R\geq g}
\left\{
\E^\mu[R]+\E^\nu[g-R]
\right\},
\end{equation*}
while the latter expresses similarly the result of optimization over a measure split,
\begin{equation*}
\sup_{0\leq\xi\leq\mu}
\left\{
\E^{\mu-\xi}[f]
+
\E^{\Sh^\nu(\xi)}[g]
\right\}
=
\sup_{R\geq g}
\left\{
\E^\mu[f\veeop R]
+
\E^\nu[g-R]
\right\}.
\end{equation*}
Another technical tool is the relaxation of stopping rules (Definition~\ref{def:relaxedrules}), 
which can be regarded either as randomization of exercise beyond the $\sigma$-algebra generated by 
the underlying process or as a partial exercise \cite{BayraktarZhou2016}.
The essential conclusion from the results of our work is that 
neither continuous nor quantized relaxation alters the 
sharp lower bound (under the same limitation of payoff convexity).

The two representations suggest complementary numerical approaches.
The convex-majorant formulation gives explicit valid subhedges at
every approximation level, while the backward shadow formulation is
sequential in the exercise dates.  They also provide natural starting
points for extensions to continuous exercise and more general payoff
structures.

\appendix

\section{Summary of terminology, notation, and standing assumptions}
\label{sec:terminology}
All measures are defined on the Borel $\sigma$-algebra ${\mathcal B}(\mathbb R)$ or ${\mathcal B}({\mathbb R}^n)$ with $n>1$, as appropriate, and are finite with finite first moment. 
The convex order 
$\alpha\cx\beta$
means that
$\E^\alpha[\varphi]\leq\E^\beta[\varphi]$
for every convex function $\varphi$ for which the expectations are
defined. We say that $\alpha$ is a submeasure of $\beta$ and write $\alpha\leq\beta$ if $\alpha(B)\leq\beta(B)$ for every Borel set $B$.
We use the notation
\begin{equation*}
\E^\mu[f]=\int_{\mathbb R}f(x)\,\mu(dx)
\end{equation*}
for the expectation of a $\mu$-integrable function $f$ where $\mu$ can be a probability measure or a submeasure of mass less than 1.  

One-dimensional marginal probability distributions $\mu_i$ are assumed to have a finite first moment and be in convex order, $\mu_1 \cx \cdots \cx \mu_N$, so at least one martingale coupling $Q$ exists. 
 We allow measures with atoms and do not assume existence of a common dominating measure. The 1-Wasserstein metric on the space of martingale couplings $Q$ is defined as 
 \begin{equation*}
 \rho_{W_1}(Q_1,Q_2) = \inf_\pi\int\! |\mathbf{x}-\mathbf{y}|\pi(d\mathbf{x},d\mathbf{y}),
 \end{equation*}
  where $|\mathbf{x}-\mathbf{y}|$ is the Euclidean metric on ${\mathbb R}^N$, $\pi$ is any coupling of measures $Q_1$ and $Q_2$ ($Q_1$ and $Q_2$ are martingale measures on ${\mathcal B}({\mathbb R}^N)$; no martingale constraint is imposed on the coupling $\pi$, a measure on ${\mathcal B}({\mathbb R}^{2N})$).  

All payoff functions $e_i$ are Borel measurable. The payoffs
$e_1,\ldots,e_N$ have no more than linear growth, and
$e_2,\ldots,e_N$ are convex.  The linear-growth assumption on $e_1$
may be replaced by $e_1\in L^1(\mu_1)$.

For a set $H\subset \mathbb{R}$, $H^c$ denotes the complement, $\mathbb{R} \setminus H$.  
For a function $f: \mathbb{R} \to \mathbb{R}$, we also write $f^c$ for the lower convex envelope but there should be no risk of confusion.

If $\alpha$ can be transported in convex order into a submeasure of
$\beta$, its shadow in $\beta$ is denoted by
$\Sh^\beta(\alpha)$.
It is the least admissible target in convex order.  

The following notation is frequently used in the paper:

\begin{itemize}
	\item $C(R_1,\ldots,R_{N-1})$ -- convex-majorant subhedge value,
	Eq.~(\ref{eq:CRi});
	
	\item $G_{\bolda}$, $G_{\bar{\bolda}}$ -- path-dependent option payoff under
	a relaxed stopping rule, Eq.~(\ref{eq:randomized-payoff});
	
	\item $\inf_Q$ -- infimum over martingale couplings with given marginals;
	
	\item $Q^{\bar{\bolda}}$ -- martingale coupling constructed for a quantized
	relaxed stopping rule $\bar{\bolda}$ in
	Sect.~\ref{sec:recursion-first};
	
	\item $\mathbf{R}=(R_1,\ldots,R_{N-1})$ -- tuple of admissible convex
	majorants, Definition~\ref{def:vcvx};
	
	\item $\sup_a$, $\sup_{\bolda}$, $\sup_{\bar{\bolda}}$ -- supremum over
	relaxed stopping rules;
	
	\item $\sup_\tau$ -- supremum over stopping times;
	
	\item $\tau_k^{\bar{\bolda}}$ -- level-crossing stopping time associated
	with a quantized relaxed rule,
	Definition~\ref{def:level-crossing-tau};
	
	\item $\tau_u^{\bolda}$ -- level-crossing stopping time associated with a
	continuous relaxed rule,
	Definition~\ref{def:level-crossing_cont_tau};
	
	\item $\tau_{\mathbf{R}}$ -- stopping time associated with convex
	majorants $\mathbf{R}$, Definition~\ref{def:tauR};
	
	\item $V(Q,H)$ -- option value under martingale coupling $Q$ and
	continuation region $H$ in the two-date problem,
	Eq.~(\ref{eq:VQH});
	
	\item $V(Q,\tau)$ -- option value under martingale coupling $Q$ and
	stopping time $\tau$, Eq.~(\ref{eq:VQtau});
	
	\item $\vqr(Q,\bolda)$ -- option value under martingale coupling $Q$
	and relaxed stopping rule $\bolda$, Eq.~(\ref{eq:vrlx_bolda});
	
	\item $V^*$ -- model-free lower value, Eq.~(\ref{eq:V});
	
	\item $\Vaff$ -- affine lower bound for the Canary option,
	Eq.~(\ref{eq:Laffine});
	
	\item $\Vcvx$ -- convex-majorant bound,
	Definition~\ref{def:vcvx} and Eq.~(\ref{eq:VNcvx});
	
	\item $\Vrlx$ -- relaxed upper bound, defined following
	Proposition~\ref{prop:continuous-to-quantized} and
	Eq.~(\ref{eq:a_abar});
	
	\item $\Vsh$ -- backward shadow bound,
	Definition~\ref{def:vshadow};
	
	\item $W_i(\mu)$ -- backward shadow value functional at exercise date
	$T_i$, Eqs.~(\ref{eq:WN}) and~(\ref{eq:shadow-recursion});
	
	\item $\mathbf{X}_i=(X_1,\ldots,X_i)$ -- history of the underlying
	process through exercise date $T_i$,
	Definition~\ref{def:relaxedrules}.
\end{itemize}

\section{Examples of shadow measures}
\label{sect:examples_shadow}
The notions of shadow measures and their associativity are central to our approach. We illustrate them with two simple examples. The first is an extremely stylized case with $\mu_1 = \delta_0$ (a unit mass at 0) and $\mu_2=U[-1,1]$ (the uniform distribution on $[-1,1]$). The second is a more realistic situation for financial applications involving normal distributions $\mu_1=N(0,1)$, $\mu_2=N(0,2)$.
\subsection{Localized mass into uniform}

\begin{figure}[htbp]
	\centering
	\includegraphics[width=0.9\textwidth]{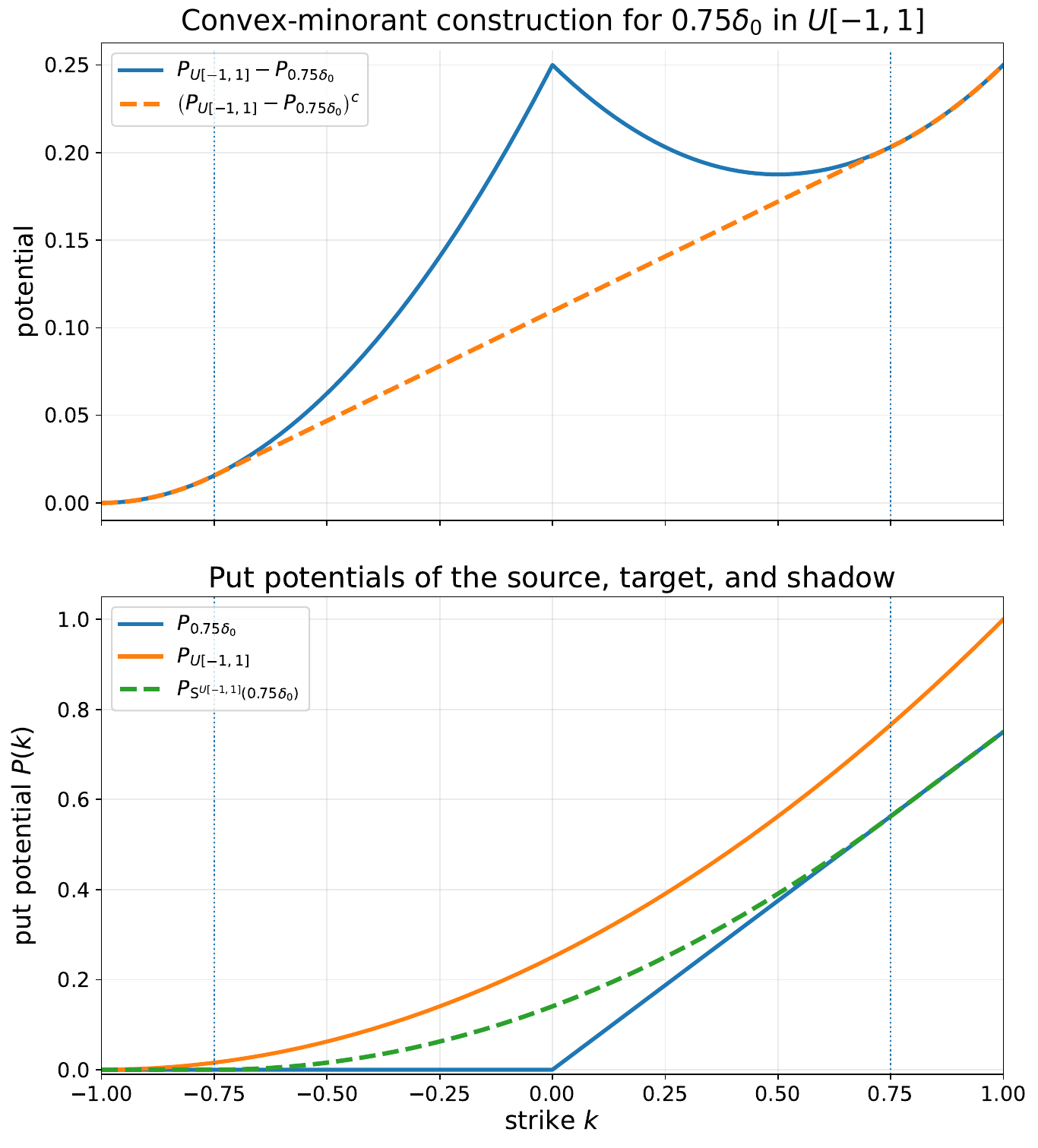}

	\caption{Put potentials for shadow of $m\delta_0$ into $U[-1,1]$ for $m=0.75$. The potentials corresponding to uniform distributions consist of segments of parabolas and straight lines.}
	\label{fig:shadow_m075}
\end{figure}

The marginals $\mu_1= \delta_0$ and $\mu_2=U[-1,1]$ are in convex order and have the same first moment. Therefore $\Sh^{\mu_2}(\mu_1)=\mu_2$. For $0<m<1$, $m\delta_0$ is a submeasure of $\mu_1$, and we have
\begin{align*}
	&\Sh^{\mu_2}(m\delta_0) = \mu_2|_{[-m,m]},\\
	&\mu_2 - \Sh^{\mu_2}(m\delta_0) = \frac{1-m}{2}(U[-1,-m]+U[m,1]).
\end{align*}
This shadow can be computed using Eq.~(\ref{eq:shadowP}). We plot the potentials corresponding to $m=0.75$ in Fig.~\ref{fig:shadow_m075}.

For $0<m_1<m_2<1$,
\begin{equation*}
	\Sh^{\mu_2 - \Sh^{\mu_2}(m_1\delta_0)}((m_2-m_1)\delta_0) = \frac{m_2-m_1}{2}(U[-m_2,-m_1]+U[m_1,m_2]),
\end{equation*}
and the associativity holds,
\begin{equation*}
	\Sh^{\mu_2}(m_1\delta_0) + \Sh^{\mu_2 - \Sh^{\mu_2}(m_1\delta_0)}((m_2-m_1)\delta_0) = \Sh^{\mu_2}(m_2\delta_0) =\mu_2|_{[-m_2,m_2]}.
\end{equation*}

\subsection{Submeasures of normal into normal}
Take $\mu_1=N(0,1)$, $\mu_2=N(0,2)$ where $N(0,\sigma^2)$ is the normal distribution with mean 0 and variance $\sigma^2$. The marginals are in the convex order and have the same first moment, therefore $\Sh^{\mu_2}(\mu_1)=\mu_2$. We take $H=(-\infty,x_0]$ and consider the shadow of $\mu_1|_H$ in $\mu_2$.

Write $\varphi$ and
$\Phi$ for the density and distribution function of the standard normal
law.  Thus,
\begin{equation*}
\mu_1(dx)=\varphi(x)\,dx,
\qquad
\mu_2(dx)=\frac{1}{\sqrt{2}}\varphi\!\left(\frac{x}{\sqrt{2}}\right)dx.
\end{equation*}

The restricted source measure has mass $\Phi(x_0)$ and first moment $-\varphi(x_0)$.
The shadow has the splice form
\begin{equation*}
{\Sh}^{\mu_2}(\mu_1|_H)
=
\mu_1|_{(-\infty,a]}
+
\mu_2|_{[a,b]},
\end{equation*}
where $a<x_0<b$ are determined by equality of mass and first moment:
\begin{align*}
	\Phi(a)
	+\Phi\!\left(\frac{b}{\sqrt{2}}\right)
	-\Phi\!\left(\frac{a}{\sqrt{2}}\right)
	&=\Phi(x_0),
	\\
	-\varphi(a)
	+\sqrt{2}\left[
	\varphi\!\left(\frac{a}{\sqrt{2}}\right)
	-\varphi\!\left(\frac{b}{\sqrt{2}}\right)
	\right]
	&=-\varphi(x_0).
\end{align*}
Solving numerically for $x_0=1$, we get
\begin{equation*}
a\approx-2.821,
\qquad
b\approx1.541.
\end{equation*}

\begin{figure}[htbp]
	\centering
	\includegraphics[width=0.9\textwidth]{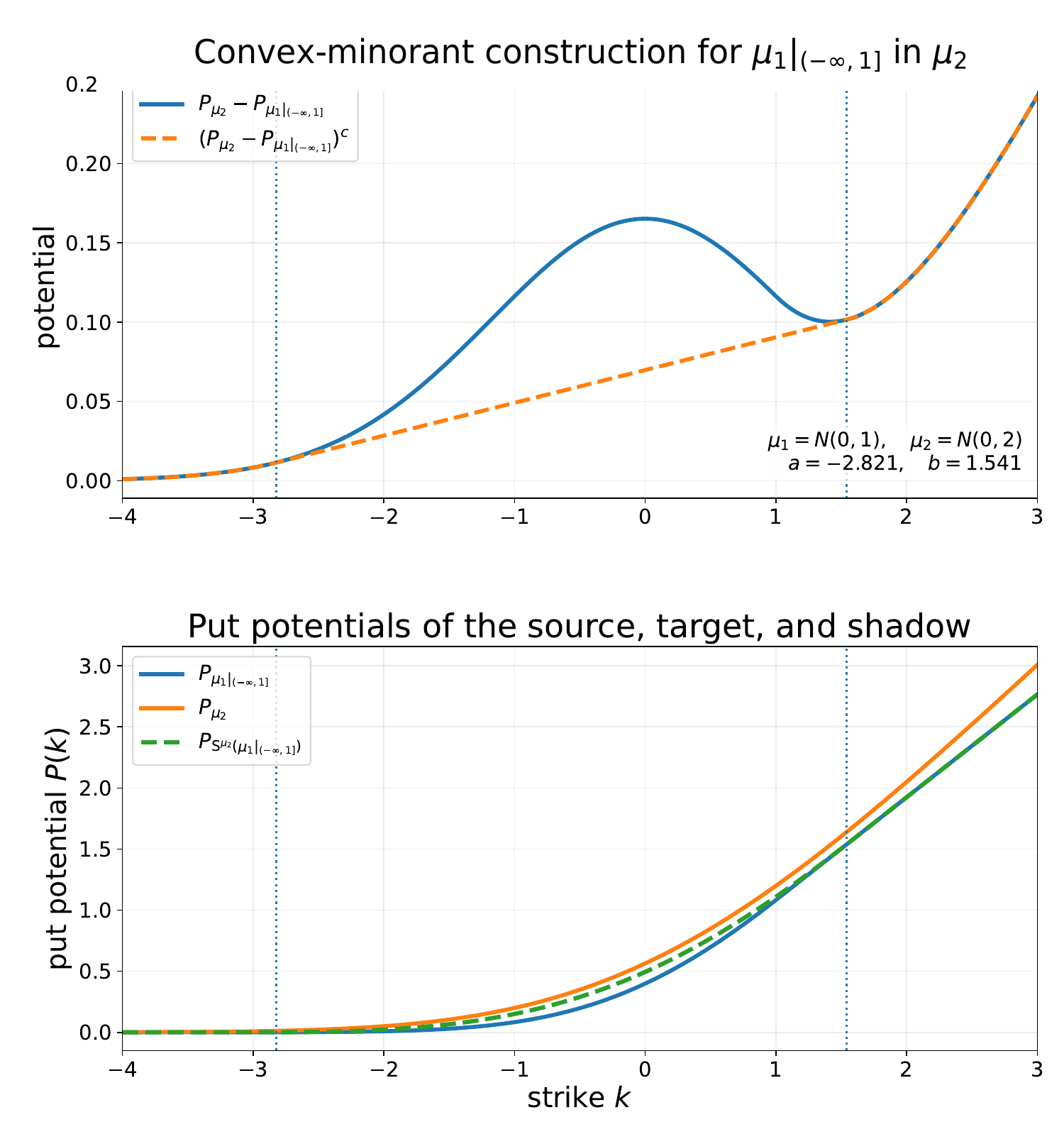}
	\caption{Put potentials for the shadow of
		$N(0,1)|_{(-\infty,x_0]}$ in $N(0,2)$ for $x_0=1$. }
	\label{fig:shadow_normal}
\end{figure}

\begin{figure}[t]
	\centering
	\includegraphics[width=0.9\textwidth]{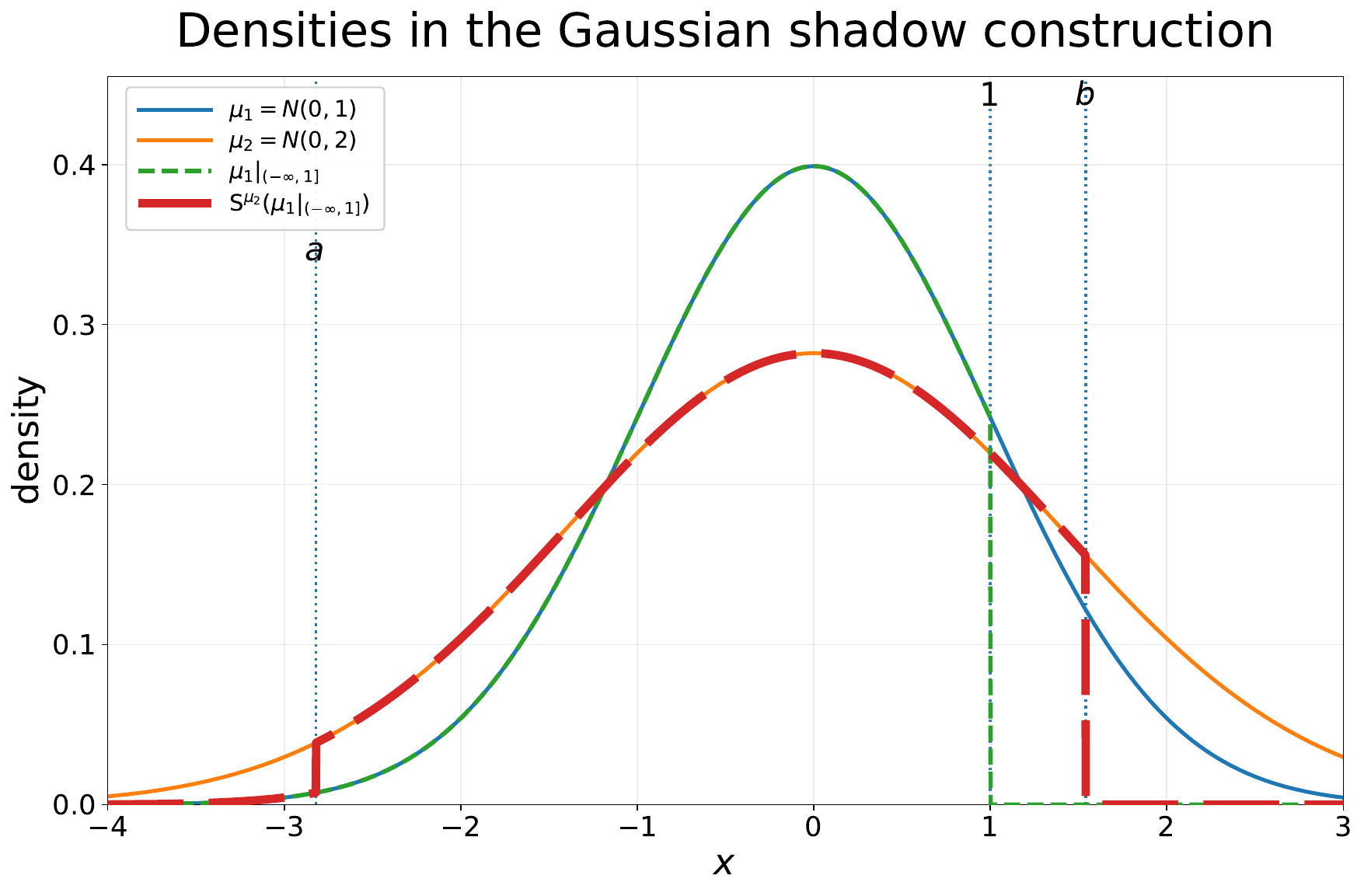}
	\caption{Densities of $\mu_1$, $\mu_2$, the restricted source
		$\mu_1|_{(-\infty,1]}$, and its shadow in $\mu_2$.  The vertical
		lines mark the points of discontinuity, $a$, $x_0=1$, and $b$.
		The shadow agrees with $\mu_1$ in the far-left tail, agrees with
		$\mu_2$ on $[a,b]$, and vanishes to the right of $b$.}
	\label{fig:gaussian-shadow-densities}
\end{figure}

We plot the potentials in Fig.~\ref{fig:shadow_normal} and the
densities in Fig.~\ref{fig:gaussian-shadow-densities}.

\clearpage
\section*{Disclaimer}

The views expressed in this paper are solely those of the author and
do not necessarily reflect the views of Bloomberg or its
affiliates.

\section*{Declaration of generative AI and AI-assisted technologies}

During the preparation of this manuscript, the author used OpenAI's
ChatGPT to assist with literature searches, editorial revision,
mathematical exposition, and the identification of points requiring
further verification. The mathematical arguments, claims, references,
and final text were reviewed and revised by the author, who takes full
responsibility for the content of the manuscript.

\clearpage
\begingroup
\small
\setlength{\parskip}{0pt}

\endgroup
\end{document}